\documentclass[amssymb, preprintnumbers, showpacs, showkeys, aps, prb, superscriptaddress, twocolumn, longbibliography,10pt]{revtex4-2}
\usepackage{color} 
\usepackage{xcolor} 
\usepackage{hyperref}
\usepackage{slashed}
\usepackage{graphicx}
\usepackage{amsmath}
\usepackage{latexsym}
\usepackage{epstopdf}
\usepackage{amsmath}
\usepackage{enumitem}
\usepackage{amssymb,amsmath}
\usepackage{multirow}
\usepackage{mathtools}
\usepackage{bbm}
\usepackage{bm}
\usepackage{amsfonts}
\usepackage{amssymb}
\usepackage{calligra}
\usepackage{calrsfs}
\usepackage{makecell}
\usepackage[normalem]{ulem}
\usepackage[USenglish,american]{babel}
\usepackage{subfigure}

\expandafter\ifx\csname package@font\endcsname\relax\else
 \expandafter\expandafter
 \expandafter\usepackage
 \expandafter\expandafter
 \expandafter{\csname package@font\endcsname}%
\fi
\begin{document}

\title{Engineering interaction potentials for stabilizing quantum quasicrystal phases}

\author{Matheus Grossklags}%
\email{matheus.grossklags@posgrad.ufsc.br}
\affiliation{Departamento de F\'\i sica, Universidade Federal de Santa Catarina, 88040-900 Florian\'opolis, Brazil}%

\author{Daniel Lima}%
\affiliation{Departamento de F\'\i sica, Universidade Federal de Santa Catarina, 88040-900 Florian\'opolis, Brazil}%

\author{Vinicius Zampronio}%
\email{v.zamproniopedroso@unifi.it}
\affiliation{Dipartimento di Fisica e Astronomia, Universit\`a di Firenze, I-50019, Sesto Fiorentino (FI), Italy}

\author{Fabio Cinti}%
\email{fabio.cinti@unifi.it}
\affiliation{Dipartimento di Fisica e Astronomia, Universit\`a di Firenze, I-50019, Sesto Fiorentino (FI), Italy}
\affiliation{INFN, Sezione di Firenze, I-50019, Sesto Fiorentino (FI), Italy}
\affiliation{Department of Physics, University of Johannesburg, P.O. Box 524, Auckland Park 2006, South Africa}

\author{Alejandro Mendoza-Coto}%
\email{alejandro.mendoza@ufsc.br}
\affiliation{Departamento de F\'\i sica, Universidade Federal de Santa Catarina, 88040-900 Florian\'opolis, Brazil}%
\affiliation{Dipartimento di Fisica e Astronomia, Universit\`a di Firenze, I-50019, Sesto Fiorentino (FI), Italy}

\begin{abstract}
    We investigate the necessary features of the pair interaction for the stabilization of self-assembled quantum quasicrystals in two-dimensional bosonic systems. Unlike the classical scenario, our results show that two-dimensional octagonal, decagonal, and dodecagonal aperiodic phases require a distinct number of properly tuned characteristic length scales for their stabilization. By using a mean field spectral variational approach and Gross-Pitaevskii numerical calculations, we determine that the dodecagonal quasicrystal structure requires at least two characteristic length scales for its stabilization, while the decagonal and octagonal patterns need at least three and four length scales, respectively. The family of pair interaction potentials considered, albeit simple, is well justified in terms of a novel experimental platform based on laser-painted interactions in a cavity QED setup. Finally, we perform a structural characterization of the quasicrystal patterns obtained and show that these phases coexist with a finite superfluid fraction, forming what can be called a super quasicrystal phase. 
\end{abstract}
\maketitle
\section{Introduction}\label{intro}
The self-assembly of complex phases such as cluster crystals~\cite{neuhaus2011phonon,cinti2014defect,de2023ultrasoft,mendoza2021ground}, quasicrystals~\cite{shechtman1984metallic,engel2007self,talapin2009quasicrystalline,fischer2011colloidal,wasio2014self,barkan2014controlled,dotera2014mosaic,pupillo2020quantum,mendoza2022exploring,grossklags2024self,mendoza2024exploring}, supersolids~\cite{cinti2010supersolid,zhang2019supersolidity,bottcher2019transient,tanzi2019observation,chomaz2019long,blakie2020supersolidity}, and topological phases~\cite{berg2007dynamical,barci2013nematic,mendoza2015nature, van2015quantum,mendoza2017quantum,mendoza2020topological,radzihovsky2020quantum} is a central topic of many-body physics, with applications in soft matter~\cite{prestipino2014hexatic,prestipino2018freezing,mendoza2021cluster,podoliak2023undulations,xia2024simple,mendoza2024melting} and hard condensed matter~\cite{bianconi2000superstripes,bianconi2001superstripes,diaz2011dynamics,fernandes2012preemptive,mendoza2012coarse,fernandes2014drives,mendoza2016modulated,mendoza2019mechanism}. In particular, quasicrystals in two and three dimensions are characterized by a lack of spatial periodicity while exhibiting long-range $n$-fold rotational symmetry (with $n>6$)  forbidden by conventional crystallography~\cite{levine1986quasicrystals,socolar1986phonons, lifshitz2011symmetry}. One singular realization of this kind of system are the so-called cluster quasicrystals~\cite{barkan2014controlled, pupillo2020quantum,mendoza2025excitations}. These systems are typically produced by isotropic soft-core pair potentials with a degenerate two minima structure in momentum space properly tuned, at moderate temperatures~\cite{zeng2004supramolecular,dotera2011quasicrystals}. To be precise, the relative position of the two competing unstable wave vectors controls the self-assembling process and hence the rotational symmetry of the stabilized quasicrystal~\cite{barkan2014controlled}. Such interplay between length scales has been extensively studied using the Lifshitz-Petrich-Gaussian model~\cite{barkan2014controlled,jiang2017stability} due to its mathematical flexibility, which allows the stabilization of periodic and aperiodic phases as the minima structure of the potential and thermal fluctuations are tuned. 

These insights have recently guided the search for novel quantum phases, showing that the combined effect of tailored interactions and quantum fluctuations can similarly stabilize quantum quasicrystals~\cite{pupillo2020quantum,mendoza2022exploring,ciardi2023quasicrystalline,grossklags2024self}. Interestingly, all studies focusing on quantum self-assembled quasicrystals analogous to the one considered by Barkan et al.~\cite{barkan2014controlled}, but at zero temperature, have been related to the stabilization of a dodecagonal phase. In this case, the literature~\cite{heinonen2019quantum,mendoza2022exploring} seems to indicate that to enhance stability or even to achieve the stabilization of these phases at all, either the higher momentum minimum should be deeper than the low momentum minimum, or extra characteristic incommensurate minima should be added to the pair interaction potential. In this scenario, to push forward the comprehension of the necessary properties of the pair interaction potential for the stabilization of quantum quasicrystals, it is essential to consider scenarios with different rotational symmetries.

Ultracold atoms provide a versatile platform for investigating quantum quasicrystals. With precise control over both internal and external degrees of freedom, these systems have been used to simulate a wide range of quantum phases~\cite{gopalakrishnan2009emergent,gopalakrishnan2010atom,vaidya2018tunable,mivehvar2021cavity,zhang2021self,karpov2022light}. However, the realization of quasicrystals~\cite{mivehvar2019emergent,mivehvar2021cavity} and other exotic phases~\cite{guo2019sign,guo2021optical} often requires long-range or oscillatory interactions, which are not typically accessible in systems of quantum gases. One novel approach involves the use of a high-finesse optical cavity and a beam shaped laser scanning the system to produce tailored effective sign-changing interactions between the atoms of the system~\cite{bonifacio2024laser}. In this work, the authors show that the effective interaction having the form $\hat{V}\left(\boldsymbol{q}\right)=g_{\text{aa}}-\Delta~\hat{\Omega}\left(\boldsymbol{q}\right)^{2}$, is controlled by the Fourier transform of the Rabi frequency profile of the laser $\hat{\Omega}\left(\boldsymbol{q}\right)$, the original contact interaction between atoms $g_{\text{aa}}$ and the parameter $\Delta$ accounting for microscopic details of the quantum mechanism involved in the interaction process. The resulting expression offers an unprecedented degree of customization, providing the same level of flexibility of the Lifshitz-Petrich-Gaussian model for the production of classical quasicrystals and other modulated phases. Although this pioneering protocol has not been tested experimentally yet, similar setups have been successfully used already to stabilize supersolids~\cite{baumann2010dicke,klinder2015dynamical}, spin textured phases~\cite{mivehvar2019cavity,guo2019sign,mivehvar2021cavity}, and extended Hubbard models~\cite{klinder2015observation,landig2016quantum}.

In the present paper, we use a mean field spectral variational approach~\cite{mendoza2022exploring,grossklags2024self,lima2025supersolid} to explore the engineering principles of the pair interaction potential in Fourier space for the stabilization of three prominent quantum quasicrystal phases with octagonal, decagonal, and dodecagonal rotational symmetries. We systematically consider the effects of having two or more competing length scales in the pair potential properly tuned in order to stabilize the corresponding quasicrystal phase and enhance its extension in the ground state phase diagram. We show that, in contrast to the classical scenario, the stabilization of quantum quasicrystals with different rotational symmetries requires pair interaction potentials with a very distinct minima structure. Next, we investigate the reasons for the much greater stability of the dodecagonal structure in comparison to the decagonal and octagonal patterns and present a geometric argument clarifying the origin of such phenomenon. Moreover, to deepen our understanding on the reported phases, we analyze the behavior of the superfluid fraction in three particular cases showing that quasicrystalline order can coexist with global superfluidity, exhibiting a phase analogous to the supersolid phase. We employed two contrasting methods to study the superfluidity, the mean field spectral  variational method and the numerical solution of the real-space Gross-Pitaevskii equation in imaginary time, showing that both techniques provide equivalent results. Finally, we discuss the results and present the concluding remarks of our work. 

The paper is organized as follows: In Section~\ref{secion2} we introduce the model Hamiltonian and the applied methodology, while Section~\ref{secion3} presents results concerning the dodecagonal structure. Decagonal quasicrystal is presented in Section~\ref{secion4}. Finally in Section~\ref{secion5} we discuss the octagonal setup.
Section~\ref{secion6} is devoted to our final remarks and conclusions.
\section{Model and Method}\label{secion2}
We consider a two-dimensional gas of $N$ bosonic particles with mass $m$ at zero temperature interacting with a pair potential $V(\boldsymbol{x})$. Within the mean field approximation, valid at high density of particles and weak enough interactions, the total energy of the system can be expressed as
\begin{equation}\label{energy_functional}
    \begin{split}
        E\left[\phi\right]=&\frac{\hbar^{2}}{2m}\int d^{2}{x} \lvert\boldsymbol{\nabla}\phi\left(\boldsymbol{x}\right)\rvert^{2} + \frac{1}{2}\int d^{2}{x} d^{2}{x}' V(\boldsymbol{x}-\boldsymbol{x}')\\
        &\times\lvert\phi\left(\boldsymbol{x}\right)\rvert^{2} \lvert\phi\left(\boldsymbol{x}'\right)\rvert^{2},
    \end{split}
\end{equation}
where $\phi\left(\boldsymbol{x}\right)$ stands for the wave function of the condensate, satisfying the normalization condition $\int d^{2}{x}\lvert\phi\left(\boldsymbol{x}\right)\rvert^{2}=N$. Moreover, we consider that the pair interaction potential is written in momentum space as $\hat{V}(\boldsymbol{q})=g_{\text{aa}}-\Delta~\hat{\Omega}\left(\boldsymbol{q}\right)^{2}$, such model of pair interaction was presented and discussed earlier in the introduction section. For a study of the stabilization conditions of distinct quasicrystal structures, we consider a rather simple laser beam model in momentum space $\hat{\Omega}\left(\boldsymbol{q}\right)$, given by the superposition of several Gaussian profiles centered at different momenta with negligible overlapping between them. As a consequence, the pair interaction potential $\hat{V}(\boldsymbol{q})$ exhibits a series of local minima properly positioned in order to favor the stabilization of a particular quasicrystalline pattern~\cite{mendoza2022exploring,grossklags2024self}. 

To stabilize a quasicrystal exhibiting $n$-fold rotational symmetry with characteristic momentum $q_{1}$, it is expected that in addition to $q_{1}$, other higher order momenta of the corresponding quasicrystal structure should also be excited. In general, if we combine two wave vectors of the extended basis $\boldsymbol{q}_{i}=q_{1}\left(\cos\left(2\pi i/n\right),\sin\left(2\pi i/n\right)\right)$ with $i=0,1,\dots,n-1$, we obtain the first group of secondary characteristic wave vectors of the corresponding pattern. For the three simplest and most relevant quasicrystalline structures, octagonal, decagonal, and dodecagonal, $n=8, 10$, and $12$, respectively, the first group of secondary wave vectors contains three or more different sets of vectors with an incommensurate length with respect to  the characteristic wave vector. Taking this into account, our approach to study the stabilization of quasicrystalline patterns involves separately exciting these length scales in $\hat{\Omega}\left(\boldsymbol{q}\right)$, thus clarifying the necessary number of minima in $\hat{V}(\boldsymbol{q})$ for the desired quasicrystal stabilization. In this way, we consider
\begin{equation}
\hat{\Omega}\left(\boldsymbol{q}\right)=\sum_{j=1}^{4}\omega_{j}\exp\left(-\frac{\left(q-q_{j}\right)^{2}}{\alpha_{j}^{2}}\right),
\end{equation}
where $\omega_{j}$, $\alpha_{j}$ and $q_{j}$ are positive parameters measuring the intensity, width and characteristic momentum of each Gaussian peak in $\hat{\Omega}\left(\boldsymbol{q}\right)$, respectively. As previously discussed, once the characteristic momentum is set by the main peak in $\hat{V}(q)$, the position of the secondary modes to be excited are located at specific irrational numbers in units of $q_{1}$.    

Considering $q_{1}$ the main wave vector of the quasicrystal structure as our unit of momentum, it is natural to choose $\lambda=1/q_{1}$ and $\epsilon=\hbar^{2}/m\lambda^{2}$ as units of length and energy, respectively. Hence, the dimensionless vector position and momentum are defined as $\boldsymbol{r}=\boldsymbol{x}/\lambda$ and $\boldsymbol{k}=\boldsymbol{q}/q_{1}$. Considering these units of energy and length, the dimensionless energy per particle functional for our model writes
\begin{equation}\label{energy_per_particle}
    \begin{split}
        \frac{E\left[\psi\right]}{N\epsilon_{0}}=&\frac{1}{2}\int\frac{d^{2}{r}}{A}\lvert\boldsymbol{\nabla}\psi\left(\boldsymbol{r}\right)\rvert^{2}+\frac{\lambda^{2}U\rho}{2}\int\frac{d^{2}{r}d^{2}{r}'}{A}\\
        &v(\boldsymbol{r}-\boldsymbol{r}')\lvert\psi\left(\boldsymbol{r}\right)\rvert^{2}\lvert\psi\left(\boldsymbol{r}'\right)\rvert^{2},
    \end{split}
\end{equation}
where $A$ stands for the area of the system and $\psi\left(\boldsymbol{r}\right)$ represents the normalized wave function, satisfying $\int_{A} d^{2}{r}\lvert\psi\left(\boldsymbol{r}\right)\rvert^{2}=A$. Furthermore, the effective pair interaction potential $U v(\boldsymbol{r})$ corresponds originally to $V(\boldsymbol{x})/\epsilon$. Here, we have introduced the dimensionless parameter $U$ to quantify the intensity of the pair interaction potential and $v(\boldsymbol{r})$ to encode the information about its spatial variation. To avoid any ambiguity in this definition we impose additionally that $\hat{v}(k=1)=-1$, without generality loss. Under these conditions, the local dimensionless density of particles is given by $\rho\left(\boldsymbol{r}\right)=\lambda^{2}\rho\lvert\psi\left(\boldsymbol{r}\right)\rvert^{2}$, where $\rho$ represents the average particles density. It is worth noticing that within our framework the dimensionless parameter $\gamma=\lambda^{2}U\rho$ controls the relative intensity of the potential energy contribution in comparison to the kinetic energy in the ground-state energy functional, or in other words the intensity of the quantum fluctuations (zero point motion) in our system. In this way, $\gamma$ naturally represents one of the running parameter of the phase diagrams for the models considered. 

To proceed, we consider that the ground-state wave function $\psi\left(\boldsymbol{r}\right)$ minimizing the energy-per-particle functional can be written in a Fourier basis. In this way, we propose
\cite{zhang2019supersolidity,mendoza2022exploring}
\begin{equation}
\label{ground_state_wave_function}
    \psi\left(\boldsymbol{r}\right)=\frac{1+\frac{1}{2}\sum_{j\neq0}c(\boldsymbol{k}_j)\cos{\left(\boldsymbol{k}_{j}\cdot\boldsymbol{r}\right)}}{\left(1+\frac{1}{4}\sum_{j\neq0}c(\boldsymbol{k}_j)^{2}\right)^{1/2}},
\end{equation}
where $c\left(\boldsymbol{k}_{j}\right)$ and $\boldsymbol{k}_{j}$ represent the Fourier amplitudes and wave vectors of the expansion. The set of wave vectors and amplitudes $\{\boldsymbol{k}_{j},c\left(\boldsymbol{k}_{j}\right)\}$ defines the type of modulated pattern under consideration and its symmetries. These wave vectors are constructed as all possible integers combinations of a vector basis  $\{\boldsymbol{e}_{j}\}$, the number of elements of the basis defines the rank of the solution. In two dimensions, periodic pattern solutions correspond to configurations of rank 2 or smaller, while quasicrystals have a rank higher than 2. In Tab.~\ref{table:basis}, we present the list of solutions considered as well as its wave vector basis, here the variational parameter $k_{m}$ sets the scale of the wave vectors lattice. In terms of possible solutions, we consider all periodic patterns typically found in quasicrystal forming systems~\cite{barkan2014controlled}: one dimensional modulations, i.e., stripes, $2$D hexagonal and square crystalline patterns, as well as two different kinds of compressed hexagonal patterns exciting simultaneously the two dominant characteristic wave vectors of the pair potential~\cite{lifshitz1997theoretical,jiang2016stability,yin2021transition} and finally, $n$-fold rotationally symmetric quasicrystals with $n=\{8,10,12\}$. Within the mean field spectral  variational analysis, the set of Fourier amplitudes $\{c_{j}\}$ and the characteristic momentum $k_{m}$ for each possible solution are determined from the minimization of the corresponding energy-per-particle functional \cite{prestipino2018clusterization,prestipino2018freezing}. Finally, the ground state is calculated selecting the solution with the lowest energy. For the special case of the homogeneous solution, all coefficients $c_{j}$ vanish except for $c_{0}=1$.
\begin{table}[!t]
    \begin{tabular}{@{}m{1.8cm}m{4.6cm}m{1.75cm}@{}}
        \cline{1-3}\hline
        Pattern & Vector basis $\{\boldsymbol{e}_{j}\}$ & Index range \\ \hline\hline
        Stripes & $k_{m}\left(1,0\right)$ &  \\
        Square & $k_{m}\left(\cos\left(2\pi j/4\right),\sin\left(2\pi j/4\right)\right)$ & $j=0,1.$ \\
        Hexagonal & $k_{m}\left(\cos\left(2\pi j/6\right),\sin\left(2\pi j/6\right)\right)$ & $j=0,1.$ \\
        Comp. Hex.1 &  $k_{m}\left(1/2,\pm\sqrt{\tilde{k}^{2}_{2}-1/4}\right)$  \\
        Comp. Hex.2 & $k_{m}\left(\tilde{k}_{2}/2,\pm\sqrt{1-\tilde{k}^{2}_{2}/4}\right)$,  \\
        8-QC & $k_{m}\left(\cos\left(2\pi j/8\right),\sin\left(2\pi j/8\right)\right)$ & $j=0,\dots,3.$ \\
        10-QC & $k_{m}\left(\cos\left(2\pi j/10\right),\sin\left(2\pi j/10\right)\right)$ & $j=0,\dots,3.$ \\
        12-QC & $k_{m}\left(\cos\left(2\pi j/12\right),\sin\left(2\pi j/12\right)\right)$ & $j=0,\dots,3.$ \\ \hline
    \end{tabular}
    \caption{Modulated patterns considered in this work and their corresponding wave vector basis $\{\boldsymbol{e}_{j}\}$. The list of of possible solutions includes: stripes, square and hexagonal crystalline patterns, compressed hexagonal patterns exciting simultaneously wave vectors on the ratio $\tilde{k}_{2}$ (see Tab.~\ref{table:secondary_wave_vectors}) and quasicrystalline patterns with eight- ($8$-QC), ten- ($10$-QC) or twelve-fold ($12$-QC) rotational symmetry.}
    \label{table:basis}
\end{table}
It is important to observe that the variational complexity of our problem can be significantly reduced by recognizing that Fourier amplitudes $c_{j}$'s corresponding to equivalent wave vectors $\boldsymbol{k}_{j}$'s after symmetry operations of the associated density pattern should be equal. For numerical purposes, the Fourier expansion in Eq.~\eqref{ground_state_wave_function} for each kind of solution is truncated using a large enough set of modes to guarantee energy convergence. Moreover, the truncation process employed guarantees that the symmetries of the corresponding pattern are fully preserved by the truncated anzats solution. 

\subsection{Pair interaction potential}
Considering the model proposed for the laser profile $\hat{\Omega}\left(\boldsymbol{q}\right)$, we represent the dimensionless pair interaction potential $\hat{v}\left(k\right)$ in the general form
\begin{equation}\label{general_pair_interaction_potential}
    \hat{v}\left(k\right)=g-\left(\sum_{j=1}^{4}d_{j}\exp\left[-\left(k-\tilde{k}_{j}\right)^{2}/\sigma_{j}^{2}\right]\right)^{2},
\end{equation}
As a consequence of our choice of length and energy scales, we have $\tilde{k}_{1}=1$ as the position of the first minimum in all cases, while the other minima in $\hat{v}\left(k\right)$ are located at selected incommensurate values favoring the stabilization of a selected $n$-fold quasicrystalline pattern, see Table~\ref{table:secondary_wave_vectors} for details.
\begin{table*}
    \centering
    \begin{tabular}{@{}m{1.5cm}>{\centering\arraybackslash}m{4.7cm}>{\centering\arraybackslash}m{4.7cm}>{\centering\arraybackslash}m{4.7cm}@{}}
        \cline{1-3}\hline
        Pattern & $\tilde{k}_{2}$ & $\tilde{k}_{3}$ & $\tilde{k}_{4}$ \\ \hline\hline
        $8$-QC & $\lvert\boldsymbol{e}_{1}+\boldsymbol{e}_{2}\rvert/\lvert\boldsymbol{e}_{1}\rvert=2\cos\left(\pi/8\right)$ & $\lvert\boldsymbol{e}_{1}-\boldsymbol{e}_{2}\rvert/\lvert\boldsymbol{e}_{1}\rvert=2\cos\left(3\pi/8\right)$ & $\lvert\boldsymbol{e}_{1}+\boldsymbol{e}_{3}\rvert/\lvert\boldsymbol{e}_{1}\rvert=2\cos\left(2\pi/8\right)$\\
        $10$-QC & $\lvert\boldsymbol{e}_{1}+\boldsymbol{e}_{2}\rvert/\lvert\boldsymbol{e}_{1}\rvert=2\cos\left(\pi/10\right)$ & $\lvert\boldsymbol{e}_{1}-\boldsymbol{e}_{2}\rvert/\lvert\boldsymbol{e}_{1}\rvert=2\cos\left(4\pi/10\right)$ & $\lvert\boldsymbol{e}_{1}+\boldsymbol{e}_{3}\rvert/\lvert\boldsymbol{e}_{1}\rvert=2\cos\left(2\pi/10\right)$\\
        $12$-QC & $\lvert\boldsymbol{e}_{1}+\boldsymbol{e}_{2}\rvert/\lvert\boldsymbol{e}_{1}\rvert=2\cos\left(\pi/12\right)$ & $\lvert\boldsymbol{e}_{1}-\boldsymbol{e}_{2}\rvert/\lvert\boldsymbol{e}_{1}\rvert=2\cos\left(5\pi/12\right)$ & $\lvert\boldsymbol{e}_{1}+\boldsymbol{e}_{4}\rvert/\lvert\boldsymbol{e}_{1}\rvert=2\cos\left(3\pi/12\right)$\\ \hline
    \end{tabular}
    \caption{Secondary wave vectors moduli excited in $\hat{v}\left(k\right)$ for each quasicrystal structure considered. The patterns $n$-QC, with $n=8$, $10$ and $12$, correspond to the octagonal, decagonal, and dodecagonal structures, respectively. The vector basis $\{\boldsymbol{e}_{j}\}$ for each kind of structure is depicted in Tab.~\ref{table:basis}.}
    \label{table:secondary_wave_vectors}
\end{table*}
To identify the key ingredients in this process, we considered pair interactions with a varying number of minima, i.e., considering cases in which some of the coefficients $d_{j}$'s are equal to zero. Once we select a particular minima structure positioned at $\{\tilde{k}_{j}\}$ in momentum space, the corresponding coefficients $\{d_{j}\}$ are obtained setting the values of the pair potential at the local minima $\{\hat{v}(\tilde{k}_{j})\}$. For all cases considered, we have $g=10$, $\sigma_{j}=0.1$ and $\hat{v}(\tilde{k}_{1})=-1$. This choice of dimensionless parameters is consistent with the range of values achievable in current experiments for the contact interaction and intensity of the laser, respectively~\cite{bonifacio2024laser}. For a systematic study, we determine the ground-state phase diagram of the different cases using as running parameters the dimensionless intensity of the pair potential $\gamma$ and the value of the pair potential at a local minimum $\hat{v}(\tilde{k}_{j})$, where $\tilde{k}_{j}$ represents a selected characteristic wave vector of the corresponding structure (see Tab.~\ref{table:secondary_wave_vectors}). The particular set of momenta $\{\tilde{k}_{j}\}$ displayed in Tab.~\ref{table:secondary_wave_vectors} corresponds to the modulus of specific higher order characteristic wave vectors of the corresponding quasicrystalline patterns. In this case, as shown in Tab.~\ref{table:secondary_wave_vectors}, we consider higher order momenta resulting from combining pairs of wave vectors of the basis (Tab.~\ref{table:basis}), for each quasicrystalline structure.

\subsection{Superfluid fraction}
To deepen our characterization of the quasicrystalline phases, we study the behavior of the superfluid fraction $\left(f_{s}\right)$ in these states. From the knowledge of the ground-state wave function $\psi\left(\boldsymbol{r}\right)$ the superfluid fraction can be estimated by means of the Legget criterion, which provides a lower bound for this quantity. Following well established literature~\cite{leggett1970can,leggett1998superfluid} we consider
\begin{equation}\label{eq:legget}
    f_{s}=\mathrm{Max}_{\alpha}\left[\int\limits_{0}^{L}\frac{dx'}{L}\left(\int\limits_{0}^{L}\frac{dy'}{L}\lvert\psi'\left(\boldsymbol{r}'\right)\rvert^{-2}\right)^{-1}\right].
\end{equation}
In Eq.~\eqref{eq:legget}, the prime symbol indicates that the coordinate system is rotated at an angle $\alpha$ with respect to the original coordinate system, consequently $\psi'\left(\boldsymbol{r}'\right)=\psi\left(x'\cos\left(\alpha\right)-y'\sin\left(\alpha\right),x'\sin\left(\alpha\right)+y'\cos\left(\alpha\right)\right)$. As occurs in the case of supersolids~\cite{blakie2024superfluid}, the angle $\alpha$ providing the highest value for $f_{s}$ coincides with the main directions of the structure, which in our case are given by the angles $\alpha_{i}=2\pi i/n$, with $i=0,...,n-1$, for an $n$-fold quasicrystalline phase. Hence, for calculation purposes, we can simply set $\alpha=0$. 

The simultaneous characterization of the superfluid properties and the density pattern allow us to identify different kinds of phases: supersolid states, when superfluidity and a crystalline density pattern coexist, super quasicrystalline states, when the quasicrystal patterns present a finite superfluid fraction, and homogeneous superfluid phases. Moreover, all these phases have their insulating counterparts when the superfluid fraction is close enough to zero~\cite{gautier2021strongly,mendoza2022exploring,grossklags2024self}.

\begin{figure*}[!t]
    \centering
    \includegraphics[width=\textwidth]{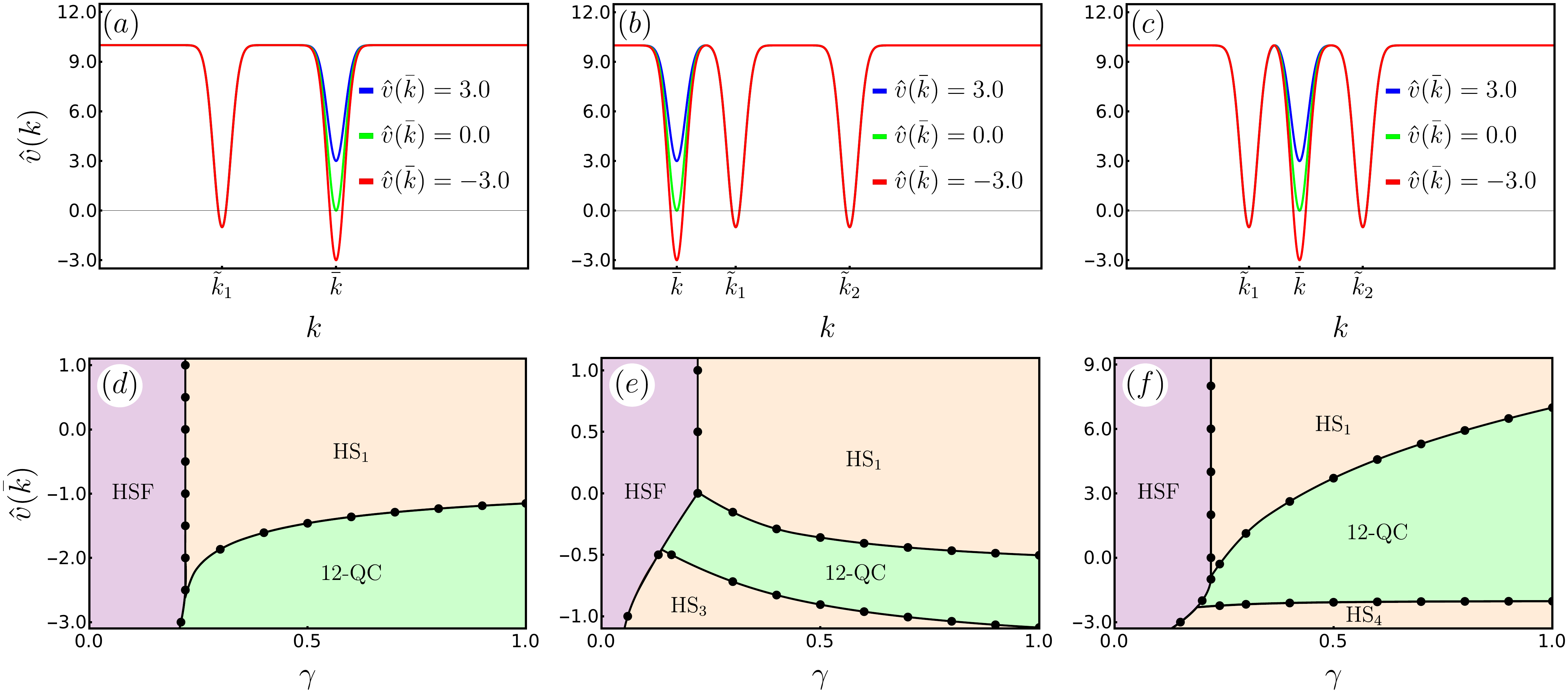}
    \caption{(a)-(c) Pair interaction potential minima structure in Fourier space for the corresponding phase diagrams in (d)-(f), where the first length scale is positioned at $k=\tilde{k}_{1}=1$ and the secondary characteristic scales are presented in Tab.~\ref{table:secondary_wave_vectors}. Phase diagram for the minima structure associated with the dodecagonal quasicrystal where (d) $\bar{k}=\tilde{k}_{2}$ (e) $\bar{k}=\tilde{k}_{3}$ and (f) $\bar{k}=\tilde{k}_{4}$ indicates the position of the minimum in reciprocal space for which we modify the depth as a parameter to determine the ground state of the system. The acronym HSF stands for homogeneous superfluid, whereas HS and $12$-QC stands for hexagonal solid and dodecagonal quasicrystal phases, respectively. The hexagonal states HS$_{i}$ refers to a hexagonal phase with characteristic wave vectors dominated by the minimum of $\hat{v}(k)$ at $\tilde{k}_{i}$.}
    \label{fig:dodecagonal_grid}
\end{figure*}

\subsection{Gross-Pitaevskii equation} 
To verify the results computed within the spectral mean field variational method, we numerically solve the Gross-Pitaevskii~\cite{pitaevskii1961vortex,gross1963hydrodynamics} equation (GPE) evolved in imaginary time, which provides the ground-state wave function and allows the calculation of the observables of our choice.

The dimensionless GPE evolved in imaginary time $\tau=it$ for the model under consideration reads 
\begin{equation}\label{eq:gpe}
    \begin{split}
        \frac{\partial\psi\left(\boldsymbol{r},\tau\right)}{\partial\tau}=&\left[\frac{1}{2}\nabla^{2}\right.\\
        &\left.-N\int d\boldsymbol{r}'v\left(\boldsymbol{r}-\boldsymbol{r}'\right)\lvert\psi\left(\boldsymbol{r}',\tau\right)\rvert^{2}\right]\psi\left(\boldsymbol{r},\tau\right).
    \end{split}
\end{equation}
Eq.~\eqref{eq:gpe} is then solved numerically using an adaptive Runge-Kutta method, with a simulation box of linear size $L=512\lambda$ and discretization $\Delta x=\Delta y=0.25\lambda$. The non-local term $v_{NL}\left(\boldsymbol{r}\right)=\int d\boldsymbol{r}'v(\boldsymbol{r}-\boldsymbol{r}')\lvert\psi\left(\boldsymbol{r}',\tau\right)\rvert^{2}$ is computed efficiently with the use of fast-Fourier transformations since ${\cal F}\left[v_{NL}\right]\left(\boldsymbol{k}\right)={\cal F}\left[v\right]\left(\boldsymbol{k}\right){\cal F}\left[\lvert\psi\rvert^{2}\right]\left(\boldsymbol{k}\right)$.

As mentioned previously, the ground-state wave function is found as the optimal stationary state of the GPE calculated using a variety of initial conditions. Our initial conditions are given by a droplet on a null background of each modulated pattern considered in our mean field spectral variational analysis placed at the center of our simulation box. When the corresponding pattern is metastable this droplet grows until covering the whole simulation box. If we integrate the GPE for long enough imaginary time a stationary state is achieved. In this way, all the metastable solutions obtained from GPE, which includes the ground state, can be compared with their counter part calculated from the variational approach. For all models considered, we observed a striking agreement between configurations computed in both ways, to show quantitatively such agreement we systematically compare the ground-state superfluid fraction using as inputs these configurations. 
\section{Dodecagonal quasicrystal}\label{secion3}
To study the stabilization of the dodecagonal quasicrystalline pattern, we begin by considering an effective interaction potential with two length scales. The first minimum in $\hat{v}(k)$ is placed at $k=\tilde{k}_{1}=1$ while the second one is positioned at $k=\tilde{k}_{2}$, which corresponds to a length scale associated with a characteristic wave vector of the dodecagonal structure (see Tab.~\ref{table:secondary_wave_vectors}). By fixing the depth of the first minimum to be $\hat{v}(\tilde{k}_{1})=-1$, it is possible to study the stabilization of the dodecagonal  phase as the vertical position of the second minimum is varied. The form of $\hat{v}(k)$ is shown in Fig.~\ref{fig:dodecagonal_grid}(a) for three distinct values of the depth of the second minimum $\hat{v}(\tilde{k}_{2})$ and the corresponding  ground-state phase diagram in the $\hat{v}(\tilde{k}_{2})$ -- $\gamma$ plane is shown in Fig.~\ref{fig:dodecagonal_grid}(d). As can be observed, three different phases are identified, a homogeneous superfluid phase (HSF) at low values of $\gamma$, a dodecagonal quasicrystal ($12$-QC) phase for low values of $\hat{v}(\tilde{k}_{2})$ and a hexagonal solid phase (HS$_{1}$) with characteristic momentum given by $\tilde{k}_{1}$. We notice that the presence of a second minimum at the first characteristic wave vector of the dodecagonal structure is sufficient to stabilize such quasicrystalline phase when this minimum is deep enough. However, in the case of a degenerate minima structure, i.e. $\hat{v}(\tilde{k}_{1})=\hat{v}(\tilde{k}_{2})$, the $12$-QC phase is only stabilized at large values of $\gamma$ in a regime where presumably particles will be largely localized, in agreement with previous works~\cite{mendoza2022exploring}. Moreover, we observe that as the value of the second minimum in $\hat{v}(k)$ is decreased, the transition from the HS$_{1}$ phase to the $12$-QC phase moves to lower values of $\gamma$. This behavior is maintained up to the point at which we observe a direct transition from the superfluid to the $12$-QC phase. Lastly, along the various phase diagrams we can identify $\gamma\approx0.23$ as the position of the homogeneous to HS$_{1}$ phase boundary. Such characteristic value can be theoretically estimated using a Ginzburg-Landau argument presented in Appendix~\ref{appendix:C}. According to this analysis, the homogeneous to modulated phase boundary is approximately located at $\gamma=-\tilde{k}^2/(4\hat{v}(\tilde{k}))$, where $\tilde{k}$ represents the characteristic wave vector of the corresponding modulated phase. A direct evaluation of this result yields $\gamma=0.25$ for the homogeneous-HS$_{1}$ phase boundary, a reasonable estimate in comparison to the actual value of the transition. Interestingly, this expression also explains the qualitative strong decreasing behavior of the $\gamma$-position of the homogeneous-modulated phase boundary for large enough values of $\vert\hat{v}(\tilde{k})\vert$, as can be observed in Fig.~\ref{fig:dodecagonal_grid}(e) and Fig.~\ref{fig:dodecagonal_grid}(f). In this region, the minima corresponding to the running parameter $\hat{v}(\tilde{k})$ is the leading one in $\hat{v}(k)$, turning $\tilde{k}$ the characteristic wave vector of the pattern.

\begin{figure}[!t]
    \centering
    \includegraphics[width=0.48\textwidth]{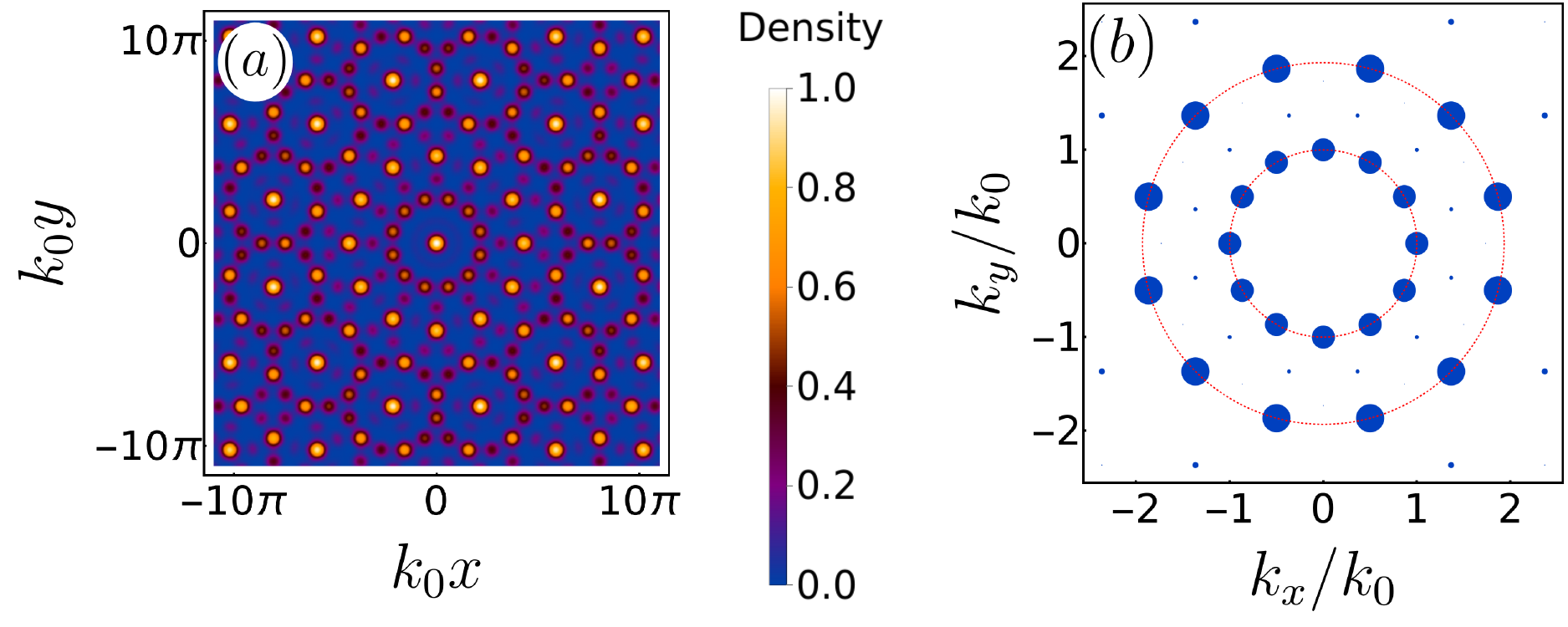}\par
    \includegraphics[width=0.48\textwidth]{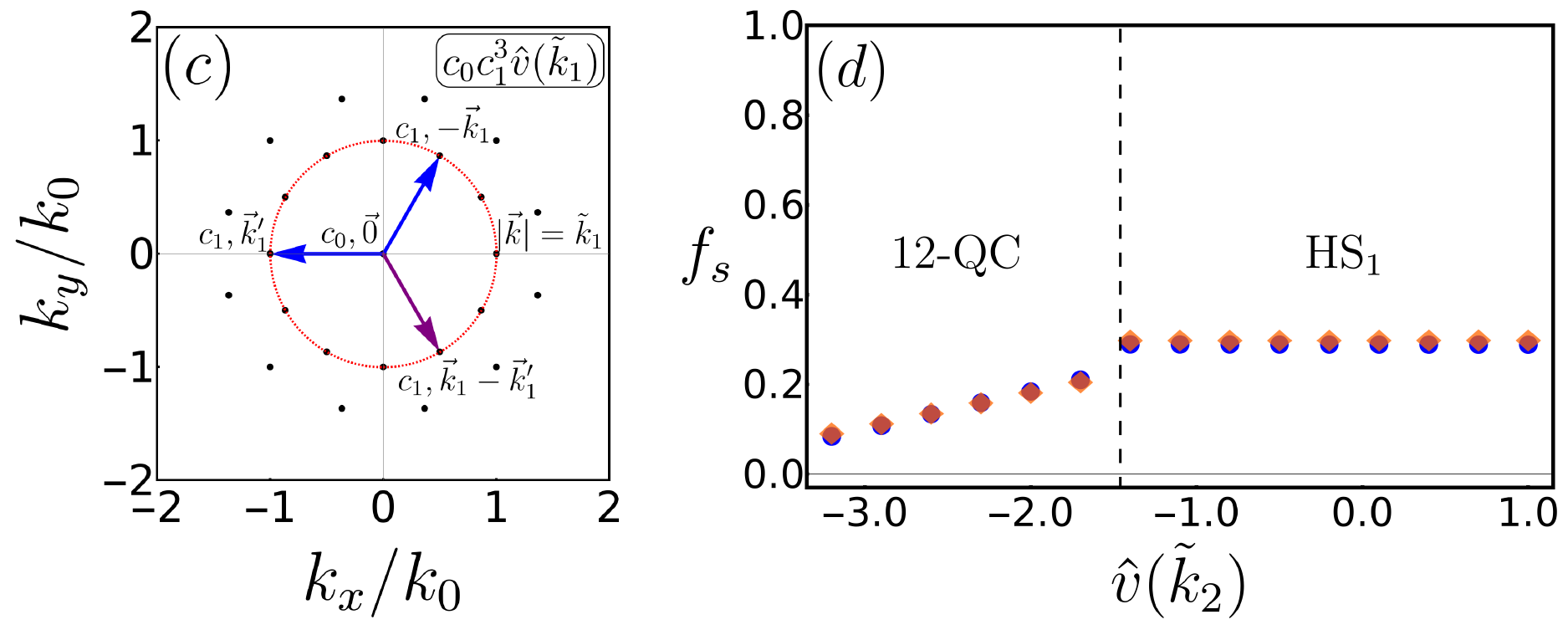}
    \caption{(a) Real space mapping for the dodecagonal structure in the phase diagram of Fig.~\ref{fig:dodecagonal_grid}(d) for parameter values $\hat{v}(\tilde{k}_{2})=-2.30$ and $\gamma=0.50$ and its corresponding (b) diffraction pattern. The red circles indicates the two minima and the size of the Fourier modes in blue disks is scaled by a linear function, where the zero harmonic mode has been omitted. (c) Most relevant energy contributions to the total energy of the quasicrystalline structures as a contraction in the $\tilde{k}_{1}=1$ layer for the dodecagonal vectors. (d) Superfluid fraction for a fixed intensity of the pair interaction, $\gamma=0.50$, as a function of $\hat{v}(\tilde{k}_2)$ for the ground state phase diagram  in Fig.~\ref{fig:dodecagonal_grid}(d). The blue points represents the superfluid fraction obtained via mean field spectral variational method while the orange diamonds are determined via the solution of the GPE.}
    \label{fig:dodecagonal_vec}
\end{figure}

To further study the stabilization of the $12$-QC phase, we modify the interaction potential including a third characteristic length scale manifesting as a minimum in $\hat{v}(k)$ at a specific wave vector of the pattern different from $\tilde{k}_{2}$. Here we decided to consider a scenario in which the minima in $\hat{v}(k)$ at $k=\tilde{k}_{1}$ and $k=\tilde{k}_{2}$ are degenerate, since this is a quite common assumption for models developing quasicrystalline phases, and investigate the effects of varying the depth of the third minimum added to $\hat{v}(k)$. Firstly, we examine the situation in which the third minima is added at $k=\tilde{k}_{3}$ (see Tab.~\ref{table:secondary_wave_vectors}), examples of the resulting pair potential $\hat{v}(k)$ are shown in Fig.~\ref{fig:dodecagonal_grid}(b). The corresponding phase diagram in the  $\hat{v}(\tilde{k}_{3})$ -- $\gamma$ plane is presented in Fig.~\ref{fig:dodecagonal_grid}(e). In agreement with the phase diagram obtained for the model with two minima, we observe that the addition of a third minimum not too deep in the pair interaction potential does not produce the stabilization of the $12$-QC phase in the range of $\gamma$ considered. However, when this minimum becomes negative we observe the existence of an interval of values in which the $12$-QC phase is again stabilized. Finally, we observe that when the third minimum added is deep enough, a reentrant transition to a new hexagonal phase (HS$_{3}$) with characteristic wave vector given by $\tilde{k}_{3}$ takes place. This behavior is already expected since an isolated dominant minimum is anticipated to stabilize a hexagonal ground state. 

\begin{figure*}[!t]
    \centering
    \includegraphics[width=\textwidth]{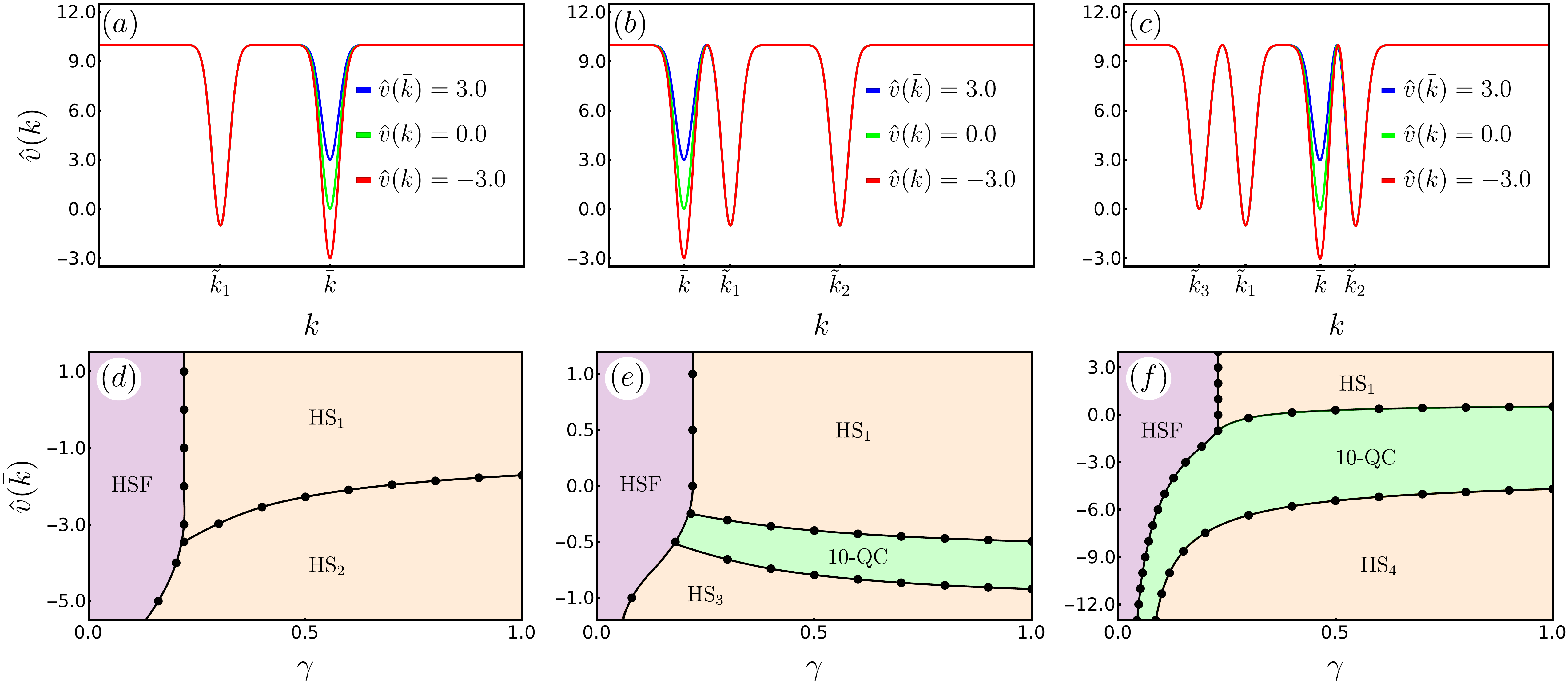}
    \caption{(a)-(c) Pair interaction potential minima structure in Fourier space for the corresponding phase diagrams in (d)-(f), in which the first length scale is positioned at $k=\tilde{k}_{1}=1$ and the secondary characteristic wave vectors are presented in Tab.~\ref{table:secondary_wave_vectors}. Phase diagram for the minima structure associated with the decagonal quasicrystal where (d) $\bar{k}=\tilde{k}_{2}$ (e) $\bar{k}=\tilde{k}_{3}$ and (f) $\bar{k}=\tilde{k}_{4}$ indicates the position of the minimum in reciprocal space for which we modify the depth as a parameter to determine the ground state of the system. The notation employed to label the phases is analogous to the one established in Fig.\ref{fig:dodecagonal_grid}.}
    \label{fig:decagonal_grid}
\end{figure*}

\begin{figure*}[!t]
    \centering
    \includegraphics[width=\textwidth]{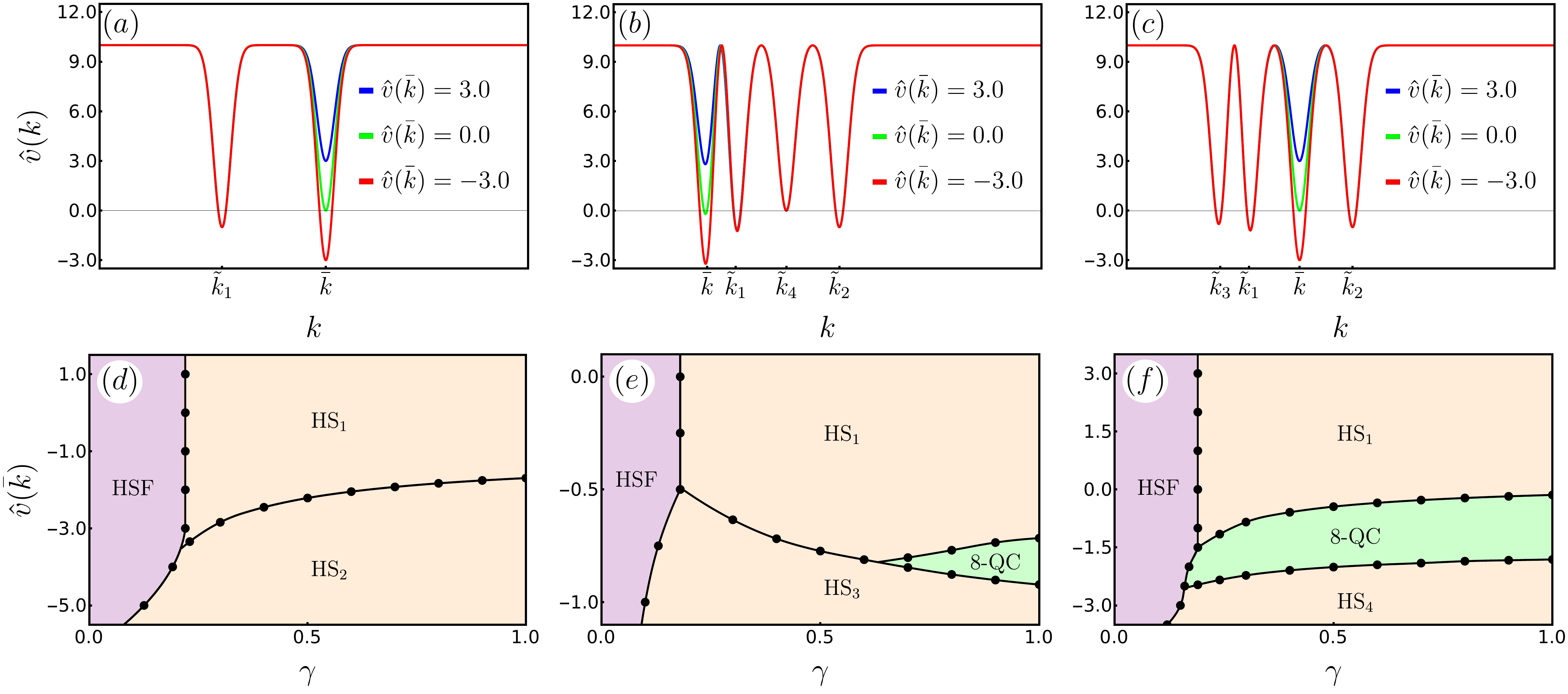}
    \caption{(a)-(c) Pair interaction potential minima structure in Fourier space for the corresponding phase diagrams in (d)-(f), where the first length scale is positioned at $k=\tilde{k}_{1}=1$ and the secondary characteristic scales are presented in Tab.~\ref{table:secondary_wave_vectors}. Phase diagram for the minima structure associated with the octagonal quasicrystal where (d) $\bar{k}=\tilde{k}_{2}$ (e) $\bar{k}=\tilde{k}_{3}$ and (f) $\bar{k}=\tilde{k}_{4}$ indicates the position of the minimum in reciprocal space for which we modify the depth as a parameter to determine the ground state of the system. The notation employed to label the phases is analogous to the one established in Fig.\ref{fig:dodecagonal_grid}.}
    \label{fig:octagonal_grid}
\end{figure*}

Lastly, we consider the excitation of a third minimum at the remaining wave vector belonging to the first generation of characteristic wave vectors of the dodecagonal quasicrystal, the one with $k=\tilde{k}_{4}$ (see Tab.~\ref{table:secondary_wave_vectors}). In this scenario, the interaction potential takes the form depicted in Fig.~\ref{fig:dodecagonal_grid}(c) and the corresponding ground-state phase diagram varying $\hat{v}(\tilde{k}_{4})$ is shown in Fig.~\ref{fig:dodecagonal_grid}(f). As can be observed, the addition of this third length scale to $\hat{v}(k)$ is much more effective in promoting the stabilization of the quasicrystalline phase when compared to the scenario shown in Fig.~\ref{fig:dodecagonal_grid}(e). In this case, even the addition of a positive minimum with a relative high value already induces the stabilization of the $12$-QC for moderate values of  $\gamma$. As expected, the strongest effect is achieved when the minimum is negative, in which case the preemptive hexagonal phase $\mathrm{HS}_1$ is not present and we have a direct transition from the HSF to the $12$-QC phase increasing $\gamma$. Moreover, as in the previous case, a further decrease of $\hat{v}(\tilde{k}_{4})$ eventually produce a transition to the HS$_{4}$ phase with characteristic wave vector close to $\tilde{k}_{4}$. It is important to note that all transitions displayed in the phase diagrams in Fig.~\ref{fig:dodecagonal_grid} are first-order. We verify this behavior by monitoring the Fourier amplitudes of the modulated phases at the homogeneous-modulated transition, which takes a significant value at the transition point. On the other hand, for the structural transitions between modulated phases, the symmetry incompatibility between quasicrystalline and the crystalline states allow us to rule out the possibility of a second-order phase transition. Moreover, we would like to mention that the scenario described above is not exclusive to the dodecagonal quasicrystal pattern, it is also observed in the case of the  decagonal (Fig.~\ref{fig:decagonal_grid}) and octagonal (Fig.~\ref{fig:octagonal_grid}) structures.

At this point, it is interesting to notice how the monotony of the phase boundaries of the dodecagonal quasicrystalline phase varies from Fig.~\ref{fig:dodecagonal_grid}(e) to (f). In the first case, the leading Fourier mode of the $12$-QC phase varies between the low and high $\gamma$ regime. At low values of $\gamma$, when $\hat{v}(\tilde{k}_{3})$ is negative but small, the leading Fourier mode of the $12$-QC is the one with the lowest value of the characteristic momenta ($\tilde{k}_3$). In this regime, the kinetic energy contribution plays a more important role than the potential one, which is a $12$-QC solution with a lower characteristic momentum is energetically preferred. Moreover, at large values of $\gamma$ the potential energy contributions dominates over the kinetic energy producing a quasicrystal pattern whose leading Fourier amplitude corresponds to the wave vector of the absolute minimum of $\hat{v}(k)$, i.e., $k=\tilde{k}_{1}$. This discussion explains the occurrence of the $12$-QC to $\mathrm{HS}_{1}$ phase transition while increasing $\gamma$. Such phase transition is driven by the system's tendency to decrease the potential energy, favoring a modulation whose characteristic wave vector coincides with the absolute minimum of $\hat{v}(k)$. The same mechanism also explains the transition from the $\mathrm{HS}_{3}$ phase to the $12$-QC phase at lower values of $\hat{v}(\tilde{k}_3)$, as we increase $\gamma$. In the case of Fig.~\ref{fig:dodecagonal_grid}(f), the situation is different. Here, the minimum at $k=\tilde{k}_{4}$ excites a set of Fourier modes acting as higher-order harmonics of the $12$-QC structure, thereby enhancing its stability. This effects remains until the minimum at $k=\tilde{k}_{4}$ becomes the dominant one and the system naturally develops a transition to a structure primarily exciting this wave vector, the $\mathrm{HS}_{4}$ phase.

Finally, we would like to point out that, in the special case shown in Fig.~\ref{fig:dodecagonal_grid}(d), the ground-state phase diagram does not exhibit a reentrant transition to the $\mathrm{HS}_{2}$ phase, even for relatively low values of $\hat{v}(\tilde{k}_{2})$, to the best of our knowledge. This interesting behavior appears to arise from the fact that, in this configuration, the two main modes of the quasicrystal ground-state wave function, corresponding to $k=\tilde{k}_{1}$ and $k=\tilde{k}_{2}$, remain active even at low values of $\hat{v}(\tilde{k}_{2})$, as both types of modes contribute to the density Fourier components at $k=\tilde{k}_{2}$. Furthermore, the simultaneous presence of these two modes enables the formation of a quasicrystalline pattern with lower kinetic energy, ultimately leading to a $12$-QC phase with lower total energy than the $\mathrm{HS}_{2}$, even in the regime of low $\hat{v}(\tilde{k}_{2})$.

The three scenarios discussed above show different routes for the stabilization of the $12$-QC phase in two dimensional bosonic gases. Although, in this case, the presence of two properly positioned minima is enough for such stabilization, differently from the classical case~\cite{barkan2014controlled,dotera2011quasicrystals}, it requires a second high-momentum minimum much deeper than the first one to stabilize the $12$-QC phase in the low $\gamma$ regime. Alternatively, the addition of a third minimum to $\hat{v}(k)$ at $\tilde{k}_{3}$ or $\tilde{k}_{4}$ enhance significantly the stability of the $12$-QC phase, even allowing us to observe this phase with a degenerate pair potential $(\hat{v}(\tilde{k}_{1})=\hat{v}(\tilde{k}_{2}))$. In this sense, by comparing Fig.~\ref{fig:dodecagonal_grid}(e) and Fig.~\ref{fig:dodecagonal_grid}(f) we conclude that in order to enhance the stability of the $12$-QC phase, it is more effective to decrease $\hat{v}(\tilde{k}_{4})$ than $\hat{v}(\tilde{k}_{3})$.

Regarding the structural properties of the $12$-QC phase, as an example, we consider the scenario presented in Fig.~\ref{fig:dodecagonal_grid}(a) with parameters $\hat{v}(\tilde{k}_2)=-2.30$ and $\gamma=0.50$. In Fig.~\ref{fig:dodecagonal_vec}(a), we show the ground-state density pattern obtained after the energy minimization, and in Fig.~\ref{fig:dodecagonal_vec}(b) a graphic of the corresponding Fourier transform of the solution. The radii of the circles are proportional to the modulus of the Fourier amplitudes, excluding the circle corresponding to the $k=0$ Fourier mode, which is not presented. The red circumferences indicate the position of the minima of $\hat{v}(k)$ in this case. As can be observed, the presence of the minima in $\hat{v}(k)$ controls which are the main modes excited in the quasicrystalline density pattern, this kind of phenomenology was verified in all scenarios considered in this work. 

To better understand the process of stabilization of the dodecagonal pattern and the greater stability of this structure in comparison with other quasicrystalline phases, as we will see later, we consider a simplified two-mode expansion including only the two main modes of the structure with $\vert\boldsymbol{k}_{j}\vert=\{\tilde{k}_{1},\tilde{k}_{2}\}$, as it is the case in Fig.~\ref{fig:dodecagonal_grid}(a). Our goal here is to assess the relative importance of the various interactions present in the energy per particle functional. Replacing Eq.~\eqref{ground_state_wave_function} into Eq.~\eqref{energy_per_particle}, we can obtain the general expression
\begin{equation}\label{potential_energy_contractions}
    \begin{split}
        \frac{E[\psi]}{N}=&\frac{1}{8Z}\sum\limits_{j}\boldsymbol{k}^{2}_{j}\lvert c\left(\boldsymbol{k}_{j}\right)\rvert^{2}+\frac{\gamma}{32Z^{2}}\sum\limits_{j,l,m}c\left(\boldsymbol{k}_{j}\right)c\left(\boldsymbol{k}_{j}+\boldsymbol{k}_{l}\right)^{\ast}\times\\
        &\times c\left(\boldsymbol{k}_{m}\right)c\left(\boldsymbol{k}_{m}-\boldsymbol{k}_{l}\right)^{\ast}\hat{v}\left(\boldsymbol{k}_{l}\right),
    \end{split}
\end{equation}
where $Z=\frac{1}{4}\sum_{j}c\left(\boldsymbol{k}_{j}\right)^{2}$ and $c\left(0\right)=2$. This expression is general and naturally remains valid for the two mode ansatz under consideration. As we can observe from Eq.~\eqref{potential_energy_contractions}, the potential energy contribution is given as a sum of contractions of four Fourier amplitudes paired carrying momentum $\boldsymbol{k}_{l}$ and $-\boldsymbol{k}_l$. At this point, it is possible to identify which contractions produce the highest contributions to the potential energy and hence the leading mechanisms of stabilization of a given quasicrystalline phase. The most relevant type of contraction is naturally the one  with the form $c\left(\boldsymbol{0}\right)c\left(\boldsymbol{k}_{1}\right)^{\ast}c\left(\boldsymbol{0}\right)c\left(-\boldsymbol{k}_{1}\right)^{\ast}\hat{v}\left(\boldsymbol{k}_{1}\right)$ and its equivalents. However, this kind of contraction is present in the energy per particle of all the solutions considered due to their pattern inversion symmetry. The next most significant contribution are of the type $c\left(\boldsymbol{0}\right)c\left(\boldsymbol{k}_{1}\right)^{\ast} c\left(\boldsymbol{k}'_{1}\right)c\left(\boldsymbol{k}'_{1}-\boldsymbol{k}_{1}\right)^{\ast}\hat{v}\left(\boldsymbol{k}_{1}\right)$, where the vectors $\boldsymbol{k}_{1}$ and $\boldsymbol{k}'_{1}$ are of equal modulus $\left(\lvert\boldsymbol{k}_{1}\rvert=\lvert\boldsymbol{k}'_{1}\rvert\right)$ and form an angle of $\pi/3$ between them. This means that the triplet of vectors $-\boldsymbol{k}_{1}$, $\boldsymbol{k}'_{1}$, and $(\boldsymbol{k}'_{1}-\boldsymbol{k}_{1})$ forms an equilateral triangle whose vector sum is zero. Hence, such a contribution to the potential energy only arises for patterns whose characteristic wave vector lattice $\left\{\boldsymbol{k}_{j}\right\}$ (see Tab.~\ref{table:basis}) exhibits $\pi/3$ rotational symmetry, i.e., crystalline patterns with hexagonal symmetry or quasicrystals possessing $6n$-fold rotational symmetry, which includes the dodecagonal pattern. Moreover, this is a contribution combining one zero-momentum Fourier amplitude with three amplitudes corresponding to vectors in the same shell, see Fig.~\ref{fig:dodecagonal_vec}(c) for a pictorial representation of the corresponding contraction. In all tests performed, this kind of contribution was responsible for around $40\%$ of the total energy of the dodecagonal pattern, which explains the greater stability of this phase in comparison to other quasicrystal solutions that do not produce such potential energy contributions.

Moreover, we investigate how the superfluid properties of the $12$-QC phase are impacted by the shape and intensity of the pair interaction potential. We estimate the superfluid fraction by means of the Legget's criterion and consider the scenario posed by Fig.~\ref{fig:dodecagonal_grid}(a). The superfluid fraction is calculated across a vertical line of the phase diagram of Fig.~\ref{fig:dodecagonal_grid}(d), specifically we consider $\gamma=0.50$ and use $\hat{v}(\tilde{k}_{2})$ as the running parameter. The results are shown in Fig.~\ref{fig:dodecagonal_vec}(d), where the blue points corresponds to the values of the superfluid fraction obtained from the density profile calculated using the mean field spectral variational method while the orange diamonds corresponds to the profile determined via the GPE solution. Within the HS$_{1}$ phase, for $\hat{v}(\tilde{k}_{2})\gtrsim-1.5$, we observe a superfluid fraction essentially constant. This is explained by the fact that Fourier modes with $\lvert\boldsymbol{k}\rvert=\tilde{k}_{2}$ are not present in the HS$_{1}$ state when the characteristic wave vector of the pattern is of the order of $\tilde{k}_{1}$. Moreover, the fact that superfluidity is significant along this phase allows us to conclude that it is actually a supersolid phase. On the other hand, in the region in which the model displays a $12$-QC phase, we can notice how the increase of the depth of the second minimum decreases the superfluid fraction. This occurs because decreasing the energy cost of the second minimum promotes higher Fourier amplitudes which implies in a more localized quasicrystalline density pattern, hindering superfluidity in the system. Nonetheless, despite the tendency of the superfluidity, within the quasicrystalline phase its value is still significant, allowing us to confirm that  this phase is actually a superfluid dodecagonal phase. Finally, the expected decreasing behavior of the superfluid fraction as a function of the pair potential intensity $\gamma$, within each modulated phase, was also confirmed. The results for each quasicrystal structure can be found in Appendix~\ref{appendix:B}.
\section{Decagonal quasicrystal}\label{secion4}
Now we proceed to study the stabilization of the decagonal quasicrystal ($10$-QC) phase. We explore systematically pair potentials with multiple length scales in the form of several minima in $\hat{v}(k)$, positioned to enhance the stability of this structure. Naturally, the first model to be considered contains two minima, the first one at $k=\tilde{k}_{1}=1$ and the second one at $k=\tilde{k}_{2}$ (see Tab.~\ref{table:secondary_wave_vectors}).  The corresponding pair interaction potential is displayed in Fig.~\ref{fig:decagonal_grid}(a), while the associated ground-state phase diagram is shown in Fig.~\ref{fig:decagonal_grid}(d). In this case, a phase transition is observed at moderate and high intensity of the pair potential $\gamma$ as the vertical position of the high momentum minima is varied. For low enough values of $\hat{v}(\tilde{k}_2)$, the system displays a transition between two hexagonal solid phases in which the characteristic wave vector of the patterns changes from $\tilde{k}_1$ (HS$_1$) to $\tilde{k}_2$ (HS$_2$). Interestingly, despite the fact that this model produces the excitation of the two main wave vectors of the decagonal quasicrystal, this pair potential configuration is still favorable to the hexagonal structure even when the high momentum minimum dominates. This indicates the necessity of introducing additional minima in $\hat{v}(k)$ to achieve the stabilization of the decagonal quasicrystal pattern.

To proceed we consider the case in which a third minimum in $k=\tilde{k}_{3}$ (see Tab.~\ref{table:secondary_wave_vectors}) is added to a two minima degenerate potential with $\hat{v}(\tilde{k}_1)=\hat{v}(\tilde{k}_2)=-1$. The resulting model for $\hat{v}(k)$ can be observed in Fig.~\ref{fig:decagonal_grid}(b) and the corresponding ground-state phase diagram using as running parameters the $\hat{v}(\tilde{k}_3)$ and $\gamma$ is shown in Fig.~\ref{fig:decagonal_grid}(e). The alternative scenario in which the third minima is added in $\hat{v}(k)$ at $k=\tilde{k}_4$ is discussed in Appendix~\ref{appendix:A}. As can be noticed, the introduction of a third characteristic length scale allows the stabilization of a 10-QC phase for a certain range of $\hat{v}(\tilde{k}_3)$. Interestingly, if $\hat{v}(\tilde{k}_3)$ is deep enough the system reenters into a hexagonal phase (HS$_{3}$) with characteristic wave vector $\tilde{k}_3$. Moreover, the quasicrystal phase exists up to the low intensity potential regime, making the system to present a direct transition from the homogeneous to the quasicrystal phase. However, this behavior is only present in a limited region of $\hat{v}(\tilde{k}_3)$, raising naturally the question about what can be done to enhance further the stability or extension of this phase within the phase diagram.

\begin{figure}[!t]
    \centering
    \includegraphics[width=0.48\textwidth]{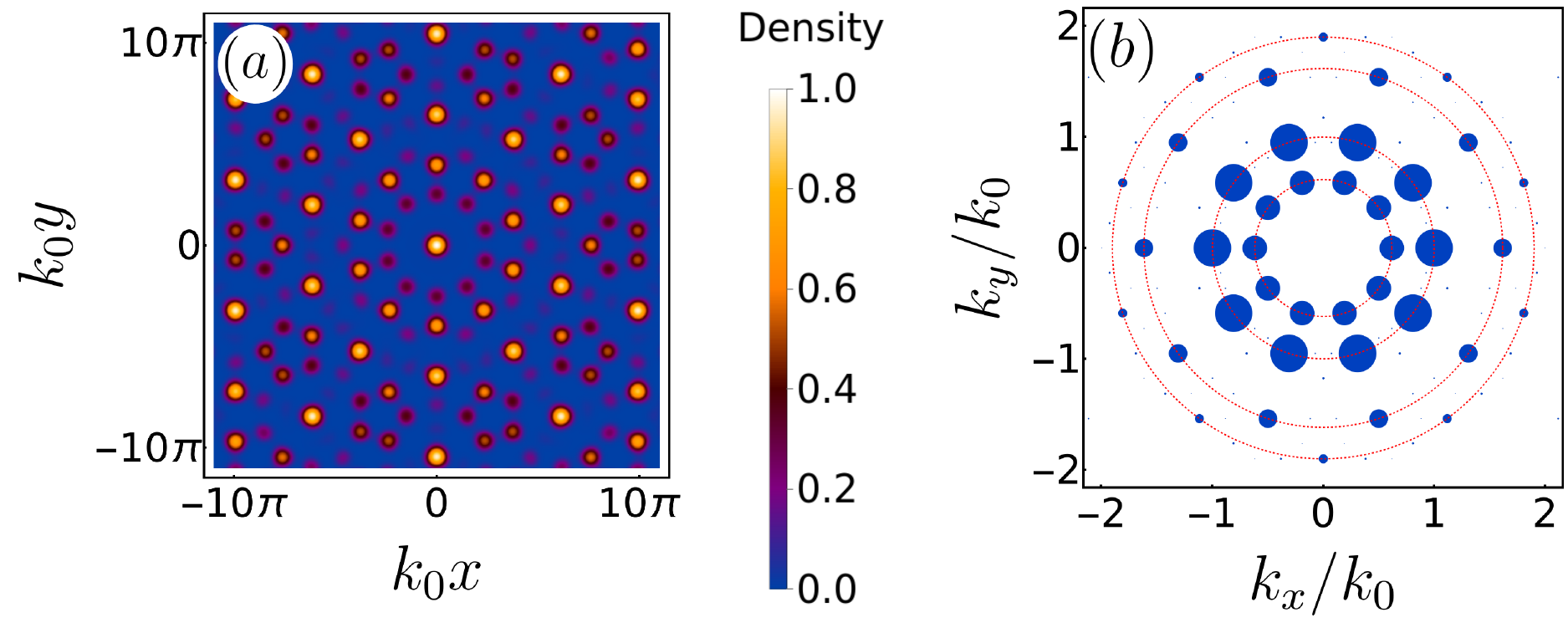}\par
    \includegraphics[width=0.48\textwidth]{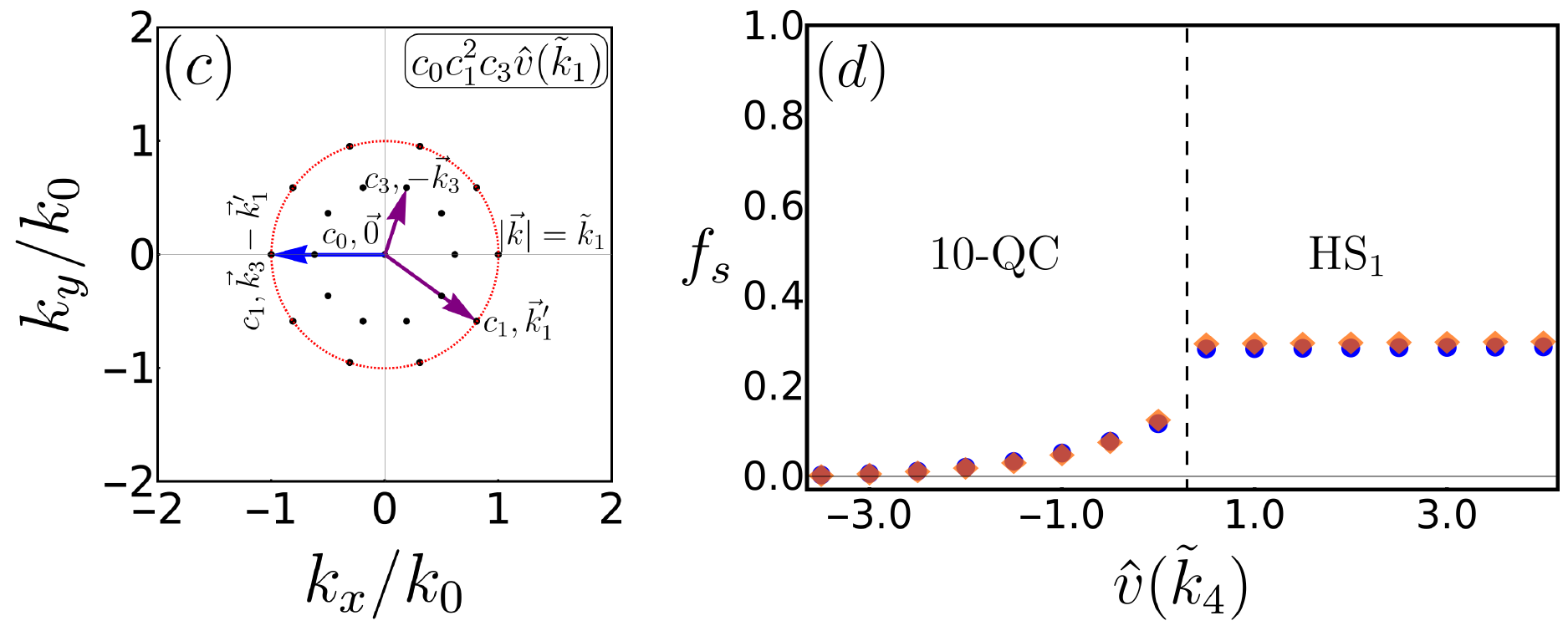}
    \caption{(a) Real space mapping for the decagonal structure in the phase diagram of Fig.~\ref{fig:decagonal_grid}(f) for parameter values $\hat{v}(\tilde{k}_{4})=-1.50$ and $\gamma=0.50$ and its corresponding (b) diffraction pattern. The red circles indicates the two minima and the size of the Fourier modes in blue disks is scaled by a linear function, where the zero harmonic mode has been omitted. (c) Most relevant energy contributions to the total energy of the quasicrystalline structures as a contraction in the $\tilde{k}_{1}=1$ layer for the decagonal vectors. (d) Superfluid fraction for a fixed intensity of the pair interaction, $\gamma=0.50$, for the the ground state phase diagram  depicted in Figure~\ref{fig:decagonal_grid}(f). The blue points represents the superfluid fraction obtained via mean field spectral variational method while the orange diamonds are determined via GPE simulations.}
    \label{fig:decagonal_vec}
\end{figure}

In order to answer this question, it is natural to consider the introduction of an additional length scale in the form of a minimum in $\hat{v}\left(k\right)$ at the remaining characteristic wave vector of the decagonal quasicrystalline structure. In this way, we add a minimum at $k=\tilde{k}_{4}$ (see Tab.~\ref{table:secondary_wave_vectors} ) to a three minima potential that does not display a quasicrystal phase at any intensity of the pair potential. In terms of minima position, this model satisfy $\hat{v}(\tilde{k}_1)=\hat{v}(\tilde{k}_2)=-1$, $\hat{v}(\tilde{k}_3)=0$ and $\hat{v}(\tilde{k}_4)$ is taken as running parameter for the construction of the corresponding phase diagram. The typical form of this kind of pair potential is displayed in Fig.~\ref{fig:decagonal_grid}(c), while in Fig.~\ref{fig:decagonal_grid}(f) we present the corresponding ground-state phase diagram obtained. As can be observed, when the depth of the fourth minimum at $\tilde{k}_4$ is increased the system transitions from a HS$_{1}$ to a $10$-QC phase. This means that only the addition of this extra minimum promotes the stabilization of the $10$-QC phase, but the extension of the quasicrystal phase is greatly enhanced. Further increasing the minimum depth leads to a transition from the decagonal quasicrystal to a hexagonal pattern with characteristic wave vector $\tilde{k}_{4}$, as expected when this minimum becomes the dominant one in the pair potential configuration.

To deepen our understanding about the $10$-QC phase, we now focus on its structural properties. In Fig.~\ref{fig:decagonal_vec}(a) and (b), we present a typical density pattern configuration and its corresponding Fourier transform for a pair potential of the type presented in Fig.~\ref{fig:decagonal_grid}(c) ($\hat{v}(\tilde{k}_{4})=-1.50$ and $\gamma=0.50$). The radius of each circle in the Fourier transform plot is proportional to the corresponding Fourier amplitude. As we can see, the main modes excited in the density pattern corresponds to those promoted by the minima structure of the pair potential as in the case of the  $12$-QC phase. An analysis of the potential energy contractions leading to dominant contributions, analogous to the one performed in the case of the dodecagonal structure, reveals that this structure does not have exclusive amplitude contractions which are specially impactful as in the previous case considered. In this case, contractions allowed for all quasicrystal structures, such as the one depicted in Fig.~\ref{fig:decagonal_vec}(c) for the decagonal pattern, are the leading contribution in the potential energy expansion. 

The behavior of the superfluid fraction for the model in Fig.~\ref{fig:decagonal_grid}(c), keeping $\gamma=0.50$ and using $\hat{v}(\tilde{k}_4)$ as our running parameter, is shown in Fig.~\ref{fig:decagonal_vec}(d). We can observe that within the region corresponding to the HS$_{1}$ phase, modifying $\hat{v}(\tilde{k}_{4})$ does not affect the superfluid fraction. As previously observed, since $\tilde{k}_{4}$ does not match any characteristic momentum of the hexagonal crystal solution with wave vector $\tilde{k}_1$, the hexagonal ground state is not affected by variations of $\hat{v}(\tilde{k}_{4})$ and hence neither the superfluid fraction. On the other hand, when the system transitions to the decagonal phase, we observe that making deeper the minimum at $\tilde{k}_4$ results in a decrease of the superfluid fraction. This behavior is a consequence of increasing the stability of the quasicrystalline phase, which results in an increase of the quasicrystal Fourier amplitudes. This in turn reflects on a more localized density profiles which hinder superfluidity. Moreover, is interesting to notice that close to the quasicrystal phase boundary, the superfluid fraction takes significant values, indicating the existence of a superfluid decagonal phase.

\section{Octagonal quasicrystal}\label{secion5}
Finally, we focus on the study of the stability of the octagonal quasicrystal ($8$-QC) phase. As in the previous cases, we initially consider a two minima model with a first minimum positioned at $\tilde{k}_{1}=1$ with $\hat{v}(\tilde{k}_{1})=-1$ and a second minimum at $\tilde{k}_{2}$ (see Tab.~\ref{table:secondary_wave_vectors}) with a varying vertical position, as it is shown in Fig.~\ref{fig:octagonal_grid}(a). The resulting phase diagram for this model is shown in Fig.~\ref{fig:octagonal_grid}(d). The obtained phase diagram presents the trivial homogeneous superfluid phase for low values of $\gamma$ and a phase transition from a hexagonal phase with characteristic wave vector $\tilde{k}_{1}$ (HS$_{1}$) to one with characteristic wave vector $\tilde{k}_{2}$ (HS$_{2}$) as $\hat{v}(\tilde{k}_2)$ is decreased. Interestingly, the addition of a third characteristic length scale of the $8$-QC phase to the pair potential $\hat{v}(k)$, either $k=\tilde{k}_{3}$ or $k=\tilde{k}_{4}$, in a scenario with two degenerate minima is not sufficient for the stabilization of the $8$-QC phase as occurs in the case of the decagonal pattern, see Fig.~\ref{fig:decagonal_vec}(e). Even in a scenario without any degeneracy we did not observe the stabilization of the $8$-QC pattern using only potentials with three characteristic length scales.

In this way we proceed by allowing the existence of a fourth minimum in $\hat{v}(k)$. Hence, our pair potential potential have local minima at $k=\tilde{k}_{1}$, $\tilde{k}_{2}$, $\tilde{k}_{3}$ and $\tilde{k}_{4}$, see Tab.~\ref{table:secondary_wave_vectors}. For this kind of model, we set $\hat{v}(\tilde{k}_{1})$ and $\hat{v}(\tilde{k}_{2})$ as degenerate minima, while we explore separately the phase diagram varying the vertical position of the other two minima present in $\hat{v}(k)$. Firstly, we set $\hat{v}(\tilde{k}_{4})=0$ and vary the depth of the minimum at $\tilde{k}_{3}$. The resulting model is presented in Fig.~\ref{fig:octagonal_grid}(b) and the corresponding phase diagram is shown in Fig.~\ref{fig:octagonal_grid}(e). Here we observe that in the low $\gamma$ regime, the system has a similar behavior to the two-minima model presented in Fig.~\ref{fig:octagonal_grid}(a), with the HS$_3$ phase having characteristic momentum $\tilde{k}_{3}$. However, for large enough potential intensity we observe the stabilization of an $8$-QC in a narrow range of $\hat{v}(\tilde{k}_{3})$. Although the $8$-QC phase is stabilized as the ground-state in this model, it only emerges in a regime of moderate to strong modulations. Taking into account that the implemented mean-field description considers a system in which particles are not localized, it is natural to seek a scenario where a direct transition from the homogeneous phase to the $8$-QC phase occurs. In such a case, one expects a significant degree of particle delocalization, thereby supporting the validity of the mean-field approximation. To this end, we now investigate an alternative configuration for the pair interaction potential, employing a different characteristic wave vector as the running parameter for our study.

In this last scenario, we continue to explore a four-minima interaction potential, but now we fix $\hat{v}(\tilde{k}_{3})=-0.6$, a value that favors the formation of the octagonal quasicrystal without turning this minimum the dominant one, see Fig.~\ref{fig:octagonal_grid}(c). Moreover, we use the depth of the minimum at $k=\tilde{k}_{4}$ as a running parameter for the construction of the phase diagram displayed in Fig.~\ref{fig:octagonal_grid}(f). As can be observed, this last family of models greatly enhances the stability of the $8$-QC phase. This is evident since, in this case, there is a significant range of $\hat{v}(\tilde{k}_4)$ for which we have a direct transition from the homogeneous to the $8$-QC phase increasing the potential intensity. Moreover, the system naturally displays a HS$_4$ phase with characteristic momentum close to $\tilde{k}_{4}$ when the potential minimum at this momentum is deep enough. The systematic study presented allow us to conclude that in order to stabilize the $8$-QC phase for the model considered, at least four characteristic wave vectors of this structure should be excited by the pair interaction potential $\hat{v}(k)$. Interestingly, the value of $\hat{v}(\tilde{k}_4)$ seem to have a much stronger impact enhancing stability of the $8$-QC than the value of $\hat{v}(\tilde{k}_3)$.   

\begin{figure}[!t]
    \centering
    \includegraphics[width=0.48\textwidth]{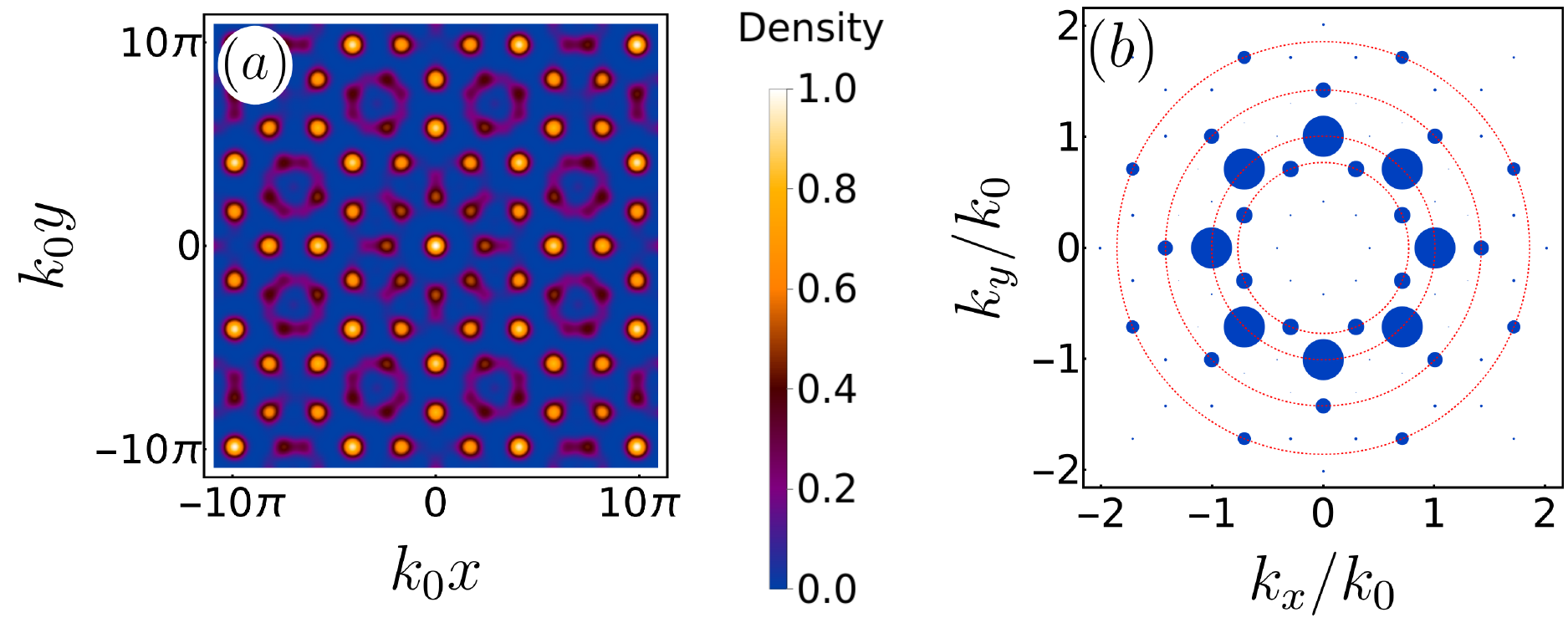}\par
    \includegraphics[width=0.48\textwidth]{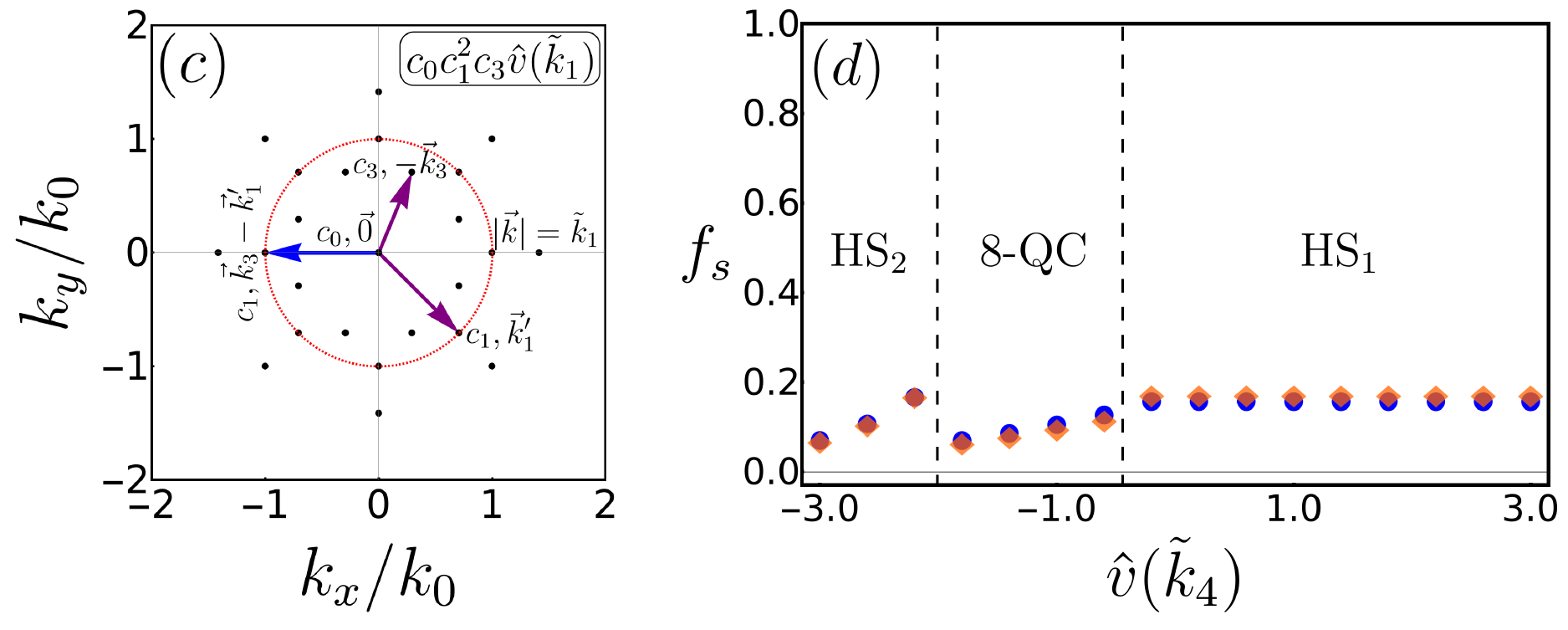}
    \caption{(a) Real space mapping for the octagonal structure in the phase diagram of Fig.~\ref{fig:octagonal_grid}(f) for parameter values $\hat{v}(\tilde{k}_{4})=-1.25$ and $\gamma=0.50$ and its corresponding (b) diffraction pattern. The red circles indicates the two minima and the size of the Fourier modes in blue disks is scaled by a linear function, where the zero harmonic mode has been omitted. (c) Most relevant energy contributions to the total energy of the quasicrystalline structures as a contraction in the $\tilde{k}_{1}=1$ layer for the octagonal vectors. (d) Superfluid fraction for a fixed value of the pair potential intensity, $\gamma=0.50$, for the ground state phase diagram in Fig.~\ref{fig:octagonal_grid}(f). The blue points represents the superfluid fraction obtained via mean field spectral variational method while the orange diamonds are determined via GPE simulations.}
    \label{fig:octagonal_vec}
\end{figure}

A real space density configuration example and its corresponding Fourier transform plot, for the model in Fig.~\ref{fig:octagonal_grid}(c), are presented in Fig.~\ref{fig:octagonal_vec}(a) and (b), respectively. The parameters employed for the variational minimization are $\hat{v}(\tilde{k}_{4})=-1.25$ and $\gamma=0.50$. It is worth noticing how the stabilization of $8$-QC phase occurs when all the unstable momenta of the engineered pair potential are excited in the density pattern of the system. Moreover, the investigation of the energy behavior using a reduced Fourier expansion for the $8$-QC reveals a scenario similar to the one observed for the $10$-QC, where no exclusive form of contraction has a high impact on the energy values and, in this context, only the introduction of additional modes and the eventual emergence of new contractions with higher momentum are effective in the stabilization of the $8$-QC phase. We show in Fig.~\ref{fig:octagonal_vec}(c) an example of the leading contractions for this phase, the specific wave vectors employed for such contractions can switch depending on which is the most unstable wave vector in the system, however preserving the rule of having two modes excited at the most unstable momenta of the system, one at the second most unstable momenta and one with zero momentum.

For completeness, we finally study the behavior of the superfluid fraction for the model described in Fig.~\ref{fig:octagonal_grid}(c) and Fig.~\ref{fig:octagonal_grid}(f), using $\hat{v}(\tilde{k}_4)$ as running parameter and fixing $\gamma=0.50$. The results obtained are displayed in Fig.~\ref{fig:octagonal_vec}(d). As previously observed, since the HS$_{1}$ phase does not excite Fourier modes with momentum $\tilde{k}_{4}$ or close to it, this phase is unaffected by changes in $\hat{v}(\tilde{k}_{4})$. In turn, when the system transition to a phase in which Fourier modes with momentum $\tilde{k}_{4}$, or close to it, are excited, to decrease the value of $\hat{v}(\tilde{k}_{4})$ promote localization of the density pattern, as explained in previous cases. This is the reason for the observed decreasing behavior of the superfluid fraction as we decrease the value of $\hat{v}(\tilde{k}_{4})$ for the 8-QC and the HS$_{4}$ phase, respectively. Moreover, we can notice that as in previous cases, also the 8-QC phase is able to host significant superfluidity confirming the existence of a superfluid octagonal  phase.\\
\section{Final Remarks}
\label{secion6}
In this work, we systematically studied the necessary conditions for the stabilization of eight-, ten-, and twelve-fold self-organized quantum quasicrystals. Unlike the classical scenario, we found that quantum quasicrystals with distinct rotational symmetries require pair interaction potentials featuring different numbers of properly selected unstable wave vectors. In the case of the dodecagonal quasicrystal, a two-minima potential with negative values is sufficient for stabilization, as previously noted in the literature~\cite{heinonen2019quantum}. Meanwhile, the decagonal and octagonal quasicrystalline structures require pair interaction potentials with at least three and four properly engineered negative minima to be stabilized, respectively.

The smaller number of unstable wave vectors required to stabilize the dodecagonal quasicrystal ground-state indicates that this structure is more robust compared to the other cases considered. As discussed in Sec.~\ref{secion3}, this enhanced stability ultimately stems from the sixfold rotational symmetry of the pattern. This symmetry allows for the presence of Fourier-mode triplets with wave vectors of equal modulus forming an equilateral triangle whose vector sum is zero. As in hexagonal crystals, this property enables cubic Fourier amplitude contributions to the potential energy, which play a significant role in reducing the total energy cost of phases exhibiting this symmetry (see Sec.~\ref{secion3}). Moreover, as can be observed in Fig.~\ref{fig:dodecagonal_grid}(f), Fig.~\ref{fig:decagonal_grid}(f), Fig.~\ref{fig:octagonal_grid}(f), as well as Appendix~\ref{appendix:B}, all quasicrystalline phases studied have a sector in which the superfluid fraction is significant, allowing us to conclude that these are actually super quasicrystal phases. These results were confirmed by extensive Gross-Pitaevskii computations, which produce equivalent configuration and superfluid fraction to those obtained with our variational spectral approach.

Although the present work considers a strictly two-dimensional model, the results obtained are also relevant for three-dimensional systems strongly confined along the $z$-direction. In such cases, the confining potential produces a particle distribution $\vert\psi_z(z)\vert^2$, often described by a Gaussian or Thomas--Fermi profile~\cite{zhang2019supersolidity,ripley2023two} along the $z$-axis. The total ground-state wave function is then typically written as $\psi_{\mathrm{3D}}(\textbf{x}) = \psi_z(z)\psi_{\mathrm{2D}}(x,y)$, where $\vert\psi_{\mathrm{2D}}(x,y)\vert^2$ describes the density pattern of the system in the $xy$-plane. This kind of ansatz enables a dimensional reduction procedure by integrating out the $z$-coordinate in the energy per particle functional, which leads to an effective two-dimensional problem in terms of $\psi_{\mathrm{2D}}(x,y)$. The resulting $2$D effective problem has essentially the same structure of the two-dimensional model considered in Eq.~(\ref{energy_per_particle}). In this case, the effective two-dimensional contact interaction corresponds to a ``renormalized'' $3$D contact interaction that depends on the width of the system along the $z$-direction.
Furthermore, regarding the laser-induced pair interaction between particles, it can be considered constant along the $z$-direction in the strong confinement regime~\cite{bonifacio2024laser}. This implies that the momentum-dependent component of the effective $2$D pair interaction will not be affected by the integration over the $z$-coordinate. Therefore, we conclude that all our results concerning the feasibility of producing quasicrystalline phases in two-dimensional bosonic systems can also be extrapolated to the more relevant experimental scenario of three-dimensional systems strongly confined along the $z$-direction.

We believe that the present results are relevant for the investigation of quasicrystal structures produced in cavity mediated interacting Bose-Einstein condensates~\cite{mivehvar2019emergent,mivehvar2021cavity,masalaeva2023tuning,orsi2024cavity}. The design on demand of cavity mediated interactions have paved a way to investigate a variety of cutting-edge many-body phases. In this regard, such systems show the interesting opportunity to operate with lattice phononic modes, a flexible manner to test quasicrystals in a quantum framework~\cite{mendoza2025excitations}. At the same time, the interplay between long-range magnetic and light-induced interactions in a Bose-Einstein condensate~\cite{mishra2023crystalline} is another promising experimental setup that, in principle, should observe 8-QC, 10-QC and 12-QC patterns displaying a finite superfluid signal. Especially relevant is also the usage of dipolar atoms in cavities~\cite{bonifacio2024laser}, allowing the implementation of many-body systems with highly tunable interactions capable of producing self-assembled quasicrystal clusters. Here, it will be of key importance to inspect the stability of octagonal as well as decagonal phases.
\section{Acknowledgments}
A.M.C. acknowledge UNIFI for financial support and hospitality.
This work was supported by the European Union through the Next Generation EU funds
through the Italian MUR National Recovery and Resilience Plan, Mission 4 Component 2–Investment 1.4–National Center for HPC, Big Data and Quantum Computing (CUP Grant No. B83C22002830001). 
F. C. and V. Z. acknowledge financial support from PNRR Ministero Universit\`a e Ricerca
Project No. PE0000023-NQSTI funded by European Union-Next-Generation EU.
The numerical solution of the GPE was performed using the XMDS2 software~\cite{dennis2013xmds2}.\\
\appendix
\section{Decagonal quasicrystal with alternative three-minima pair interaction potential}
\label{appendix:A}

\begin{figure}[!h]
    \centering
    \includegraphics[width=1.0\linewidth]{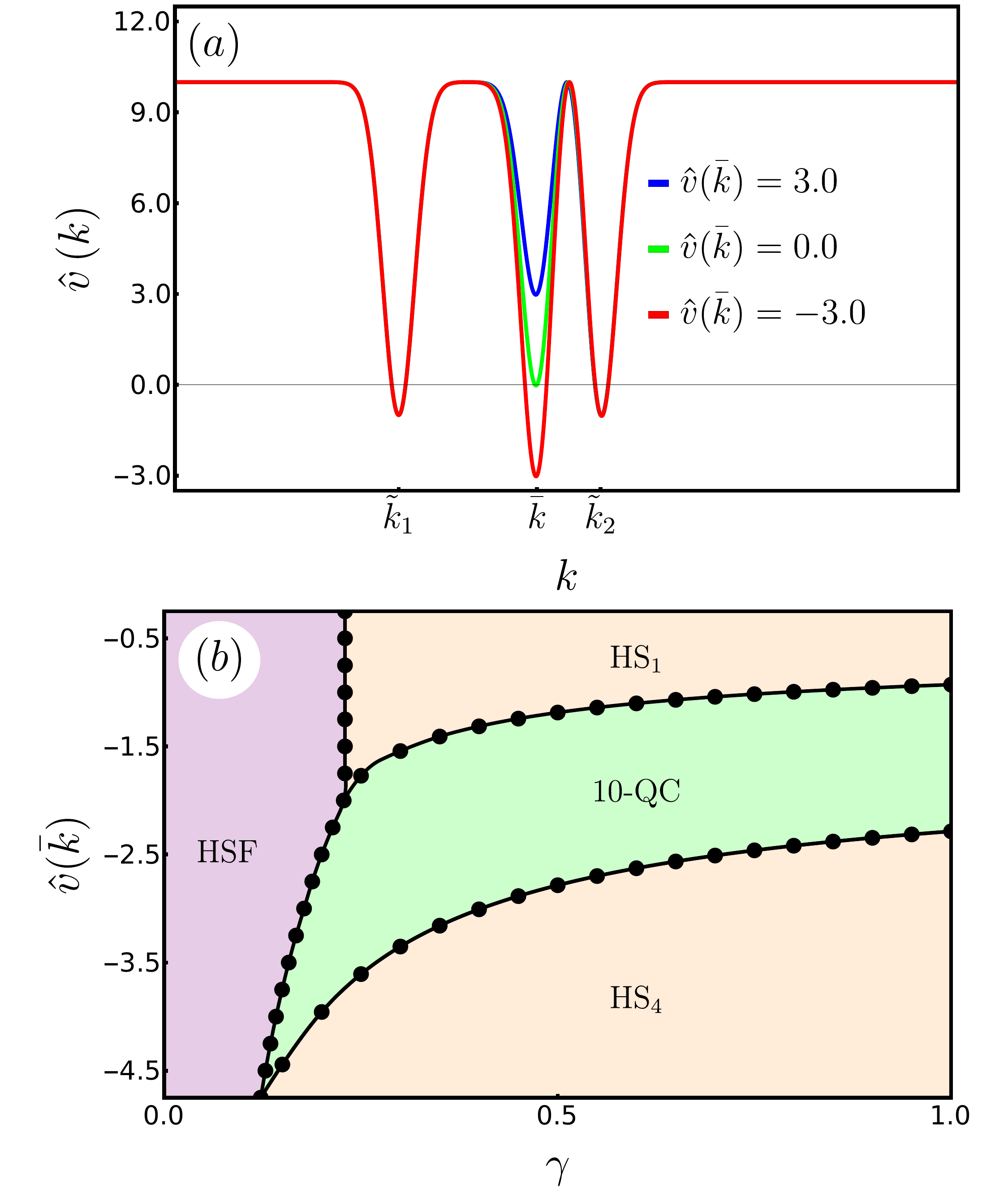}
    \caption{(a) Pair interaction potential containing a three minima structure in Fourier space. The minima are located at the wave vectors $\tilde{k}_{1}$, $\tilde{k}_{2}$ and $\tilde{k}_{4}$ for the decagonal structure (see Tab.~\ref{table:secondary_wave_vectors}). (b) Ground-state phase diagram of the model using $\hat{v}(\bar{k})$, with $\bar{k}=\tilde{k}_{4}$ and $\gamma$ as running parameters. The acronyms HSF, $\mathrm{HS}_{i}$ and $10$-QC stand for homogeneous superfluid, hexagonal solid and decagonal quasicrystal phase, respectively. The sub-index ``i'' in the HS acronym indicate a phase with characteristic wave vector $\tilde{k}_{i}$ as defined in Tab.~\ref{table:secondary_wave_vectors}.}
    \label{fig:decagonal_appendix_model}
\end{figure}

An alternative scenario to the one considered in Fig.~\ref{fig:decagonal_grid}(e) corresponds to the case in which a third minimum is added at $k=\tilde{k}_{4}$ (see Tab.~\ref{table:secondary_wave_vectors}). To make possible the comparison with the phase diagrams in Fig.~\ref{fig:decagonal_grid}, we consider the situation where $\hat{v}(\tilde{k}_{1})=\hat{v}(\tilde{k}_{2})=-1$. In this way, the resulting pair interaction potential in momentum space, $\hat{v}(k)$, is shown in Fig.~\ref{fig:decagonal_appendix_model}(a). The corresponding phase diagram, using $\hat{v}(\tilde{k}_{4})$ and $\gamma$ as running parameters, is shown in Fig.~\ref{fig:decagonal_appendix_model}(b). As can be observed, for sufficiently high values of $\hat{v}(\bar{k})$, the hexagonal phase with characteristic wave vector $\tilde{k}_{1}$ is the most stable modulated phase. As $\hat{v}(\bar{k})$ decreases, the system undergoes a transition to a decagonal quasicrystal ($10$-QC) phase and eventually to an $\mathrm{HS}_{4}$ phase in the low $\hat{v}(\bar{k})$ regime. Additionally, for small values of $\gamma$, the system exhibits a transition to a homogeneous superfluid (HSF) phase, as expected.


By comparing the phase diagram obtained for the present model with that of Fig.~\ref{fig:decagonal_grid}(e), we observe that the extension of the $10$-QC phase along the $\hat{v}(\bar{k})$ axis is increased in the present case. Moreover, the $10$-QC region in the phase diagram of Fig.~\ref{fig:decagonal_grid}(e) is shifted toward higher values of $\hat{v}(\bar{k})$. This shift results from the lower momentum of the third minimum excited by the pair potential in that case, which generates density Fourier modes with lower kinetic energy, thus favoring the early stabilization of the $10$-QC phase. Finally, we conclude that, from a general perspective, the study of the model shown in Fig.~\ref{fig:decagonal_appendix_model} confirm the necessity of the competition among three length scales to stabilize the $10$-QC phase.
\section{Superfluid fraction as a function of $\gamma$}\label{appendix:B}
In this appendix, we discuss the behavior of the superfluid fraction as $\gamma$ is varied across various models that display quasicrystalline phases. This study considers three representative examples, one for each type of quasicrystal. Specifically, we compute the superfluid fraction for a fixed pair potential $\hat{v}(k)$, using $\gamma$ as the running parameter. For our analysis, we have selected the models presented in Fig.~\ref{fig:dodecagonal_grid}(a), Fig.~\ref{fig:decagonal_grid}(c), and Fig.~\ref{fig:octagonal_grid}(c), corresponding to the dodecagonal, decagonal, and octagonal quasicrystals, respectively. As in the main text, two independent methods are employed for this calculation: direct GPE simulations and the spectral variational method.

\begin{figure}[!t]
    \centering
    \includegraphics[width=\linewidth]{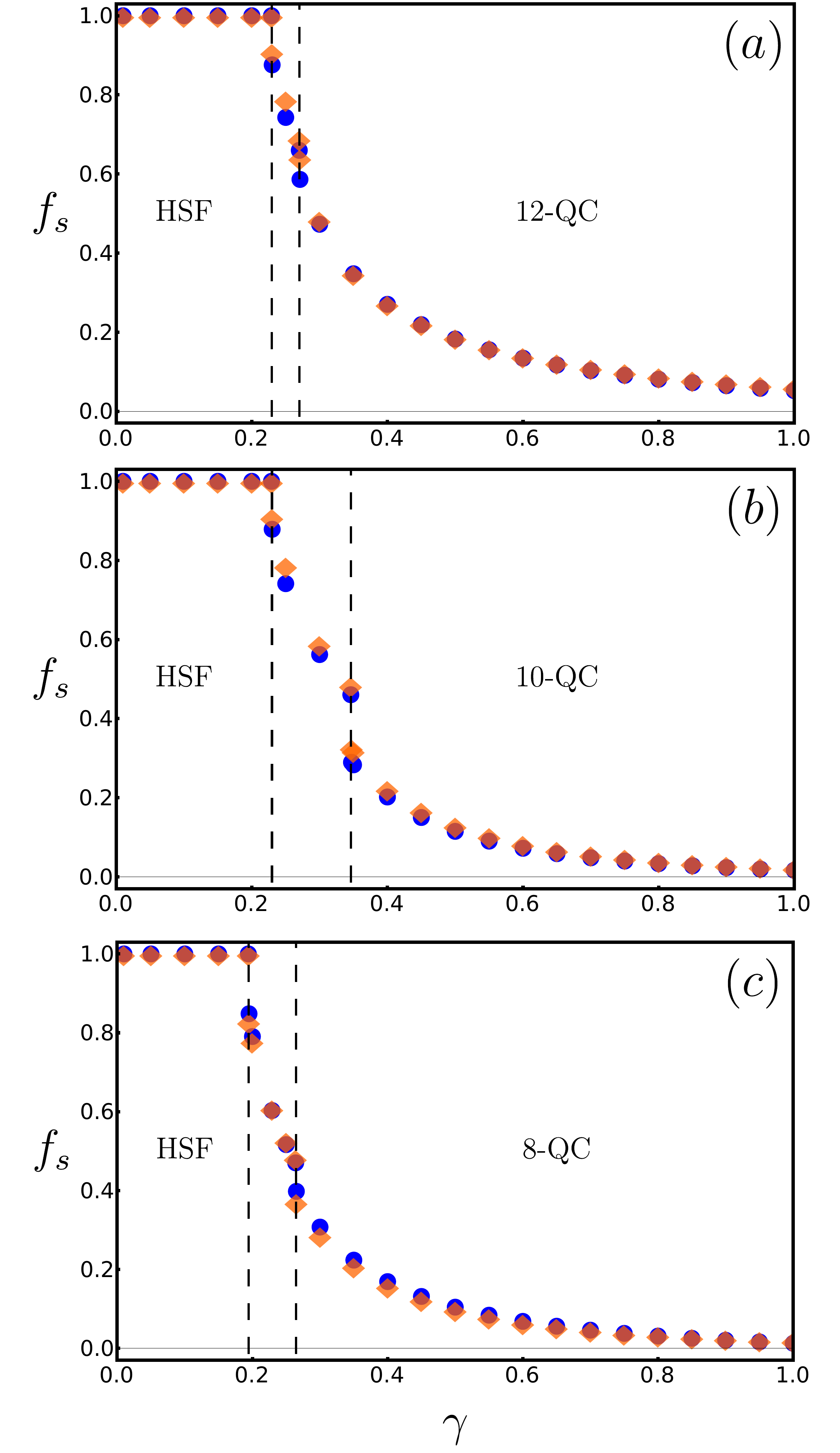}
    \caption{Superfluid fraction $(f_{s})$ as a function of the pair potential intensity $\gamma$ for the dodecagonal ($12$-QC), decagonal ($10$-QC) and octagonal ($8$-QC) quasicrystal structures. (a) Model presented in Fig.~\ref{fig:dodecagonal_grid}(a), considering $\hat{v}(\tilde{k}_2)=-2.0$. (b) Model in Fig.~\ref{fig:decagonal_grid}(c) for $\hat{v}(\tilde{k}_4)=0.0$. (c) Model from  Fig.~\ref{fig:octagonal_grid}(c) setting $\hat{v}(\tilde{k}_4)=-1.0$. Black dashed lines indicate the phase boundaries while the intermediate phase in all diagrams corresponds to the HS$_{1}$ phase. The blue points corresponds to the results obtained via spectral mean-field method while orange diamonds represent values from GPE simulations.}
    \label{fig:superfluid_fraction_density}
\end{figure}

In Fig.~\ref{fig:superfluid_fraction_density}, we present the resulting superfluid fractions as a function of $\gamma$, considering $\hat{v}(\tilde{k}_{2})=-2.0$ in panel (a), $\hat{v}(\tilde{k}_{4})=0.0$ in panel (b), and $\hat{v}(\tilde{k}_{4})=-1.0$ in panel (c). As can be observed, within a given modulated phase, the superfluid fraction generally decreases as $\gamma$ increases. This behavior is a direct consequence of the growing modulation amplitude and increased particle localization as the interaction strength, controlled by $\gamma$, becomes stronger. In addition to this monotonic trend, discontinuities in the superfluid fraction are visible at the phase boundaries, as expected in a first-order phase transition scenario.

In the regime of small to moderate values of $\gamma$, the system simultaneously exhibits spatial order and superfluidity in both periodic and aperiodic phases, thereby confirming the existence of supersolid and super quasicrystalline phases. In the large $\gamma$ regime, the superfluid fraction approaches zero; however, it never strictly vanishes. This is a well-known limitation of the mean-field variational description of Bose-Einstein condensates. Finally, we would like to remark the excellent agreement between the results obtained via the two methods employed, confirming the reliability and consistency of the superfluid fraction calculations.

\section{Homogeneous–hexagonal phase boundary behavior}\label{appendix:C}

The behavior of the homogeneous–hexagonal ($\mathrm{HS}_{1}$) phase boundary can be approximately understood using a Ginzburg–Landau approach in terms of the leading Fourier mode amplitude ($c_1$) of the hexagonal pattern. This description is naturally valid in the regime of weak modulations, hopefully occurring in the vicinity of the homogeneous–hexagonal phase boundary.

In this case, the ground state wave function can be written by the following ansatz:
\begin{eqnarray}
\nonumber
\psi(\textbf{x}) &=& \frac{1}{Z}\left[1 + c_1\left(\cos(kx) + \cos\left(-\frac{kx}{2} + \frac{\sqrt{3}ky}{2}\right)\right.\right. \\
&+& \left.\left. \cos\left(-\frac{kx}{2} - \frac{\sqrt{3}ky}{2}\right)\right)\right],
\end{eqnarray}
where the normalization factor is given as $Z = \sqrt{1 + \frac{3c_1^2}{2}}$. By inserting this ansatz into the energy per particle functional of Eq.~(\ref{energy_per_particle}) and expanding the resulting expression in powers of $c_1$, we obtain, up to order $O(c_1^5)$:
\begin{equation}
    \begin{split}
        E_{\mathrm{hex}}(c_{1},k) =\,& E_{\mathrm{hom}} 
        + \left(\frac{3\tilde{k}_{1}^{2}}{4} + 3\gamma\hat{v}(\tilde{k}_{1})\right)c_{1}^{2} 
        + \alpha_{3}c_{1}^{3} \\
        & + \alpha_{4}c_{1}^{4} + \dots,
    \end{split}
\end{equation}
where $\alpha_{i}$ are the coefficients of the power-series expansion in $c_1$, and $E_{\mathrm{hom}} = \gamma\hat{v}(0)/2$ denotes the energy per particle of the homogeneous phase.

Although the presence of a cubic term indicates the occurrence of a first-order transition, a good estimate of the transition point can be obtained from the instability threshold condition $
3\tilde{k}_{1}^{2}/4 + 3\gamma\hat{v}(\tilde{k}_{1}) \leq 0$. This yields an approximate boundary located at
\begin{equation}
    \gamma = -\frac{\tilde{k}_{1}^{2}}{4\hat{v}(\tilde{k}_{1})}.
\end{equation}

In all cases considered, the main Fourier component of the $\mathrm{HS}_{1}$ phase corresponds to $\tilde{k}_{1} = 1$, while the value of $\hat{v}(\tilde{k}_{1})$ is fixed to $-1$. Thus, the position of the phase boundary is estimated as $\gamma = 0.25$ for all cases. This observation explains why the position of the homogeneous to $\mathrm{HS}_{1}$ transition is essentially constant for all phase diagrams. Moreover, it is important to mention that the position for the phase boundary evaluated here is an estimate based on the leading Fourier harmonic expansion for the hexagonal pattern. The actual value observed in the phase diagrams is slightly lower due to the relevant role of higher order Fourier harmonics in the wave function expansion.

\bibliography{ref} 
\end{document}